\documentclass{aa}  

\usepackage{graphicx}
\usepackage[varg]{txfonts}
\usepackage{subfigure}
\usepackage{amssymb}
\usepackage{cancel}
\usepackage{bm}
\usepackage{hyperref}
\usepackage{color}
\usepackage{footnote}
\newcommand{\rev}[1]{{#1}}

\defcitealias{2015IBVS.6126....1A}{A15}
\defcitealias{2001ApJ...556..265W}{WG01}

\begin{document}

   \title{Violent environment of the inner disk of RW Aur A\\probed by the 2010 and 2015 dimming events\thanks{This work is based on observations made with the ESO telescopes at the Paranal Observatory under programmes ID 294.C-5047 and 382.C-0711}}
   
\titlerunning{Violent environment of the inner disk of RW Aur A}

   \author{S. Facchini\inst{1,}\inst{2}   
           \and
           C.~F. Manara\inst{3}
           \and
           P.~C. Schneider\inst{3}
           \and
           C.~J. Clarke\inst{2} 
           \and
           J. Bouvier\inst{4} 
           \and
           G. Rosotti\inst{2}
           \and
           R. Booth\inst{2}
           \and
           T.~J. Haworth\inst{2,}\inst{5}
          }

   \institute{Max-Planck-Institut f\"ur Extraterrestrische Physik, Giessenbachstrasse 1, 85748 Garching, Germany,\\\email{facchini@mpe.mpg.de}    
                         \and
                 Institute of Astronomy, Madingley Rd, Cambridge, CB3 0HA, UK
                 \and
                        Scientific Support Office, Directorate of Science, European Space Research and Technology Centre (ESA/ESTEC), Keplerlaan 1, 2201 AZ Noordwijk, The Netherlands
                \and
                        UJF-Grenoble 1/CNRS-INSU, Institut de Plan\`{e}tologie et d'Astrophysique de Grenoble (IPAG) UMR 5274, 38041 Grenoble, France
                \and
                Astrophysics Group, Imperial College London, Blackett Laboratory, Prince Consort Road, London SW7 2AZ, UK
             }

   \date{Received; accepted}

 
  \abstract
  {}   
   {The young binary system RW Aur shows strong signatures of a recent tidal encounter between the circumprimary disk and the secondary star. The primary star has recently undergone two major dimming events ($\Delta$mag $\approx2$ in $V$ band) whose origin is still under debate. To shed light on the mechanism leading to the dimming events, we study the extinction properties, accretion variability, and gas kinematics using absorption lines from the material that is obscuring RW Aur A.}
   {We compared our moderate-resolution X-Shooter spectra of the dim state of RW Aur A with other spectral observations. In particular, we analysed archival high-resolution UVES spectra obtained during the bright state of the system to track the evolution of the spectral properties across the second dimming event. Since the X-Shooter spectrum is flux calibrated, we provide new synthetic photometry of RW Aur A during the dim state.}
   {The spectrum obtained during the dim state shows narrow absorption lines in the Na and K optical doublets, where the former is saturated. With a velocity of $-60\,$km/s, these lines indicate that during the dim state the disk wind is either enhanced or significantly displaced into the line of sight. The photometric evolution across the dimming event shows a grey extinction, and is correlated with a significant reduction of the EW of all photospheric lines. Emission lines that trace accretion do not vary significantly across the dimming.}
   {From comparing our observations with complementary results from the last years, we conclude that the dimming event is related to a major perturbation on the inner disk. We suggest that the inner disk is occulting (most of) the star and thus its photosphere, but does not occult the accretion regions within a few stellar radii. Since observations of the outer disk indicate that the disk is modestly inclined ($45$ - $60^\circ$), we propose that the inner disk might be warped by an as yet unseen (sub-) stellar companion, which may also explain the $2.77$\,day periodic variability of the spectral lines.}

   \keywords{accretion, accretion disks -- planetary systems: protoplanetary disks -- binaries: general -- stars: individual: RW Aurigae -- stars: variables: T Tauri, Herbig Ae/Be}

   \maketitle
%
\section{Introduction}

\label{sec:intro}

The binary system RW Aur consists of two K-type stars \citep[with masses of $1.4$ and $0.9M_\odot$,][]{1997ApJ...490..353G,2001A&A...376..982W}
and has a significantly high ratio of the accretion rate onto the primary star A \citep[$2-10\times10^{-7}M_\odot/$yr,][]{1995ApJ...452..736H} to the circumprimary disk mass \citep[$\sim 10^{-3}M_\odot$,][]{2005ApJ...631.1134A}. The system is also known to possess a powerful jet launched from star A \citep[e.g.][]{2003A&A...405L...1L}. The secondary star B is located $\approx1.5\arcsec$ away from star A \citep[e.g.][]{2006A&A...452..897C}, and the whole system is located at $140\,$pc from Earth \citep{2007A&A...474..653V}.
\citet{2006A&A...452..897C} observed a huge tidal tail trailing the primary disk in $^{12}$CO rotational emission lines, and they interpreted it as the dynamic signature of a tidal interaction between the circumprimary disk and the secondary star B. This interpretation has been supported by detailed hydrodynamical modelling \citep{2015MNRAS.449.1996D}, which places the system into the rare category of protoplanetary disk systems observed to be undergoing a dynamical encounter. The gravitational disturbance might be the cause of the high accretion rate, since tidal encounters can remove angular momentum from the inner regions of the disk \citep[e.g.][]{1993MNRAS.261..190C,1994ApJ...424..292O}.

RW Aur A has recently undergone two major dimming events \citep[][]{2013AJ....146..112R,2015A&A...577A..73P}. The first one lasted for $\sim180$ days, with a dimming of $1.5-2$ magnitudes in the $V$ band \citep{2013AJ....146..112R}. The second dimming event started before October 2014 \citep[and ended in August 2016][]{2016MNRAS.463.4459B}, with an even deeper occultation ($\Delta V\gtrsim2$\,mag). Photometric studies have shown that during the second dimming event the extinction is nearly grey from optical \citep[][hereafter \citetalias{2015IBVS.6126....1A}]{2015IBVS.6126....1A} to near-infrared (NIR) bands \citep[][]{2015A&A...584L...9S}, suggesting that the occulting material has undergone significant grain growth. Moreover, \citet{2015A&A...584L...9S} have detected an enhancement in the gas column density along the line of sight to star A during the second dimming event ($N_{\rm H}\approx2\times10^{22}$\,cm$^{-2}$) by comparing X-ray observations of RW Aur A during bright and dim states. Finally, \citet[][]{2015IBVS.6143....1S} have shown that the second dimming event is correlated to an excess in the NIR (in $L$ and $M$ bands), and they proposed that this excess might be due to an increased emission from hot dust ($\sim1000$\,K) at $\sim0.1$\,AU, suggesting that the occulting material is associated with the inner regions of the disk. \rev{Archival WISE data have confirmed the same observational result \citep{2016MNRAS.463.4459B}.}

To explain the observed dimming events, different models have been invoked. Using hydrodynamical modelling, \citet{2015MNRAS.449.1996D} have shown that a tidal bridge between the two stars can obscure RW Aur A and that the column density might be sufficient to explain the dimming \citep[as initially proposed by][]{2013AJ....146..112R}. \citet[][]{2015A&A...577A..73P} have suggested that the extinction could be due to an outburst of a stellar wind entrapping dust grains of the inner region of the disk. Finally, \citet{2015A&A...584L...9S} have proposed that the observations could be explained by geometric variations in the inner disk, similarly to the AA Tau system \citep[e.g.][]{2013A&A...557A..77B}, or possibly in the V1184 Tau system \citep[e.g.][]{2016A&A...588A..20G}. These scenarios are more thoroughly described in Sect. \ref{sec:discussion}.

An insight into the kinematic properties of the occulting material and the accretion rate evolution of star A is essential if we are to determine the physical origin of the dimming event and to distinguish between the possible interpretations reported above. In Sect. 2 of this paper we present new spectral data obtained during the second dimming event with the X-Shooter spectrograph mounted on the Very Large Telescope (VLT), together with an analysis of archival VLT/UVES data taken during the system bright state. In Sect. \ref{sec:results} we analyse the observational data and compare the two sets of spectra. In Sect. \ref{sec:discussion} we discuss the possible interpretations of our and recent observations, and in Sect. \ref{sec:conclusions} we summarise our conclusions.

\section{Observations and data reduction}
\label{sec:observations}

\subsection{X-Shooter observations}
We have obtained new spectra of RW Aur A using the VLT/X-Shooter instrument \citep{2011A&A...536A.105V} during the night of March 19, 2015 (DDT Pr.Id. 294.C-5047, PI Facchini). At the time of observation the seeing conditions were very good (0.6\arcsec at zenith, $\sim$0.86\arcsec\  at the airmass of the target). A set of large slits (1.6x11\arcsec-1.5x11\arcsec-1.2x11\arcsec\  in the UVB, VIS, and NIR arms) and a set of narrow slits (0.5x11\arcsec\  in the UVB arm, 0.4x11\arcsec\  in the VIS and NIR arms) were used for the acquisition of the data. The spectra obtained using the large slits have lower spectral resolution but do not suffer slit losses, and are thus absolutely flux calibrated. Conversely, the spectra taken with the narrow slits lead to the highest spectral resolution (R$\sim$18000 at $\lambda$$\sim$600 nm), but must be corrected in flux using the data taken with the large slits. The observing strategy we adopted consisted of a short exposure (70s, 45s, and 45s in each arm) in stare mode with the large slit, followed by a complete ABBA nodding cycle with the narrow slits with exposures 160s, 70s, and 80s in each arm at any nodding position, and finally another short stare exposure (50s, 30s, and 35s in each arm) with the large slit again. This procedure was repeated twice. We emphasise that star B lay out of the slit in all our observations.

\begin{figure*}
\begin{center}
\includegraphics[width=0.9\textwidth]{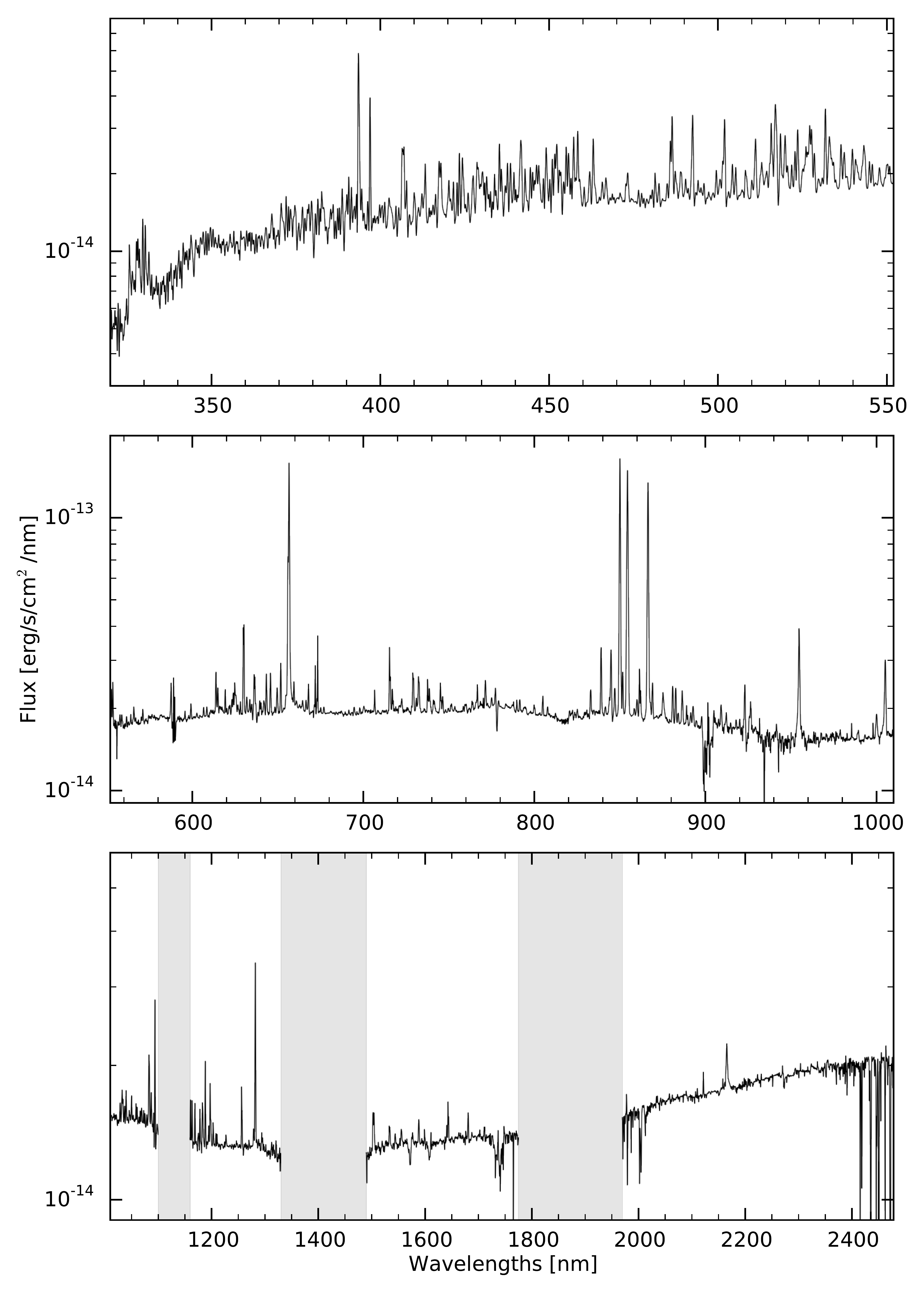}
\end{center}
\caption{X-Shooter flux-calibrated spectrum of RW Aur A after telluric correction. The grey regions are dominated by atmospheric absorption.}
\label{fig:spectrum}
\end{figure*}

The spectra were reduced using the standard ESO pipeline for X-Shooter \citep{2010SPIE.7737E..28M} version 2.5.2 with the same procedure as in \citet{2013A&A...558A.114M}. The reduction was carried out independently for each arm, and the flux calibration in the pipeline was performed using the flux standard star observed during the night. Each 1D extracted spectrum obtained with the pipeline was corrected for telluric using a standard telluric spectrum obtained at similar airmass conditions as the target and with slits with the same width. Then, the two narrow-slit spectra taken successively were co-added to enhance the signal-to-noise ratio (S/N). Finally, the combined narrow-slit spectra were rescaled to the large-slit spectra to correct for slit losses. The final spectra cover the wavelength region from $\lambda\sim$320 nm to $\lambda\sim$2500 nm with a resolution R$\sim$10000 of up to $\lambda\sim$550 nm and from $\lambda\sim$ 1000 nm, and R$\sim$18200 in between. The entire X-Shooter spectrum is shown in Fig. \ref{fig:spectrum}.

\subsection{UVES observations}

In the next sections we compare spectral features of the dimmed state from the X-Shooter spectrum with previous higher resolution UVES \citep{2000SPIE.4008..534D} spectra of RW Aur A obtained during its bright state between 2008 and 2010 (Pr.Id. 382.C-0711, PI Whelan). The dates of the UVES observations are reported in Table \ref{tab:uves}. These data were obtained using 0.6\arcsec\ wide slits, leading to a resolution R$\sim$70000. The spectra are divided into three arms, the blue arm covering the wavelength range $\lambda\lambda$ 376-498 nm, and the red arms covering the range from $\lambda\sim$498 nm to $\lambda\sim$700 nm when the central wavelength was set to 600 nm, or from $\lambda\sim$670 nm to $\lambda\sim$1042 nm with central wavelength 860 nm. In both cases a gap of $\sim$8 nm centred on the central wavelength is present in the red arm spectra. 

The reduction was carried out with the standard ESO pipeline for UVES version 5.7.0. The output of the pipeline is the extracted 1D spectrum, which is also flux calibrated in the blue arm and in the red arm when the central wavelength is 860 nm. However, this flux calibration is obtained using the standard response curve and does not account for slit losses, so  the absolute flux calibration may be incorrect. We therefore subsequently only use the shape of the emission and absorption lines from the UVES data, not the fluxes. Finally, no correction for telluric lines was performed on these UVES spectra.

\begin{table}
\centering
\begin{tabular}{lll}
\hline
Instrument & Colour in fig. & Date (d/m/y)\\
\hline
\hline
UVES    & cyan          & 07/12/2008 \\
UVES    & green         & 11/01/2009 \\
UVES    & orange                & 18/01/2009 \\
UVES    & violet                & 25/01/2009 \\
UVES    & magenta       & 08/02/2009 \\
UVES    & pink          & 16/11/2009 \\
UVES    & blue          & 28/12/2009 \\
UVES    & red           & 03/01/2010 \\
X-Shooter               & black         & 19/03/2015 \\
\hline
\label{tab:uves}
\end{tabular}
\caption{Dates of the UVES and X-Shooter spectra analysed in this work. The colour coding used in the figures is also reported.}
\end{table}

\section{Results}
\label{sec:results}

\begin{figure*}
\begin{center}
\includegraphics[width=\columnwidth]{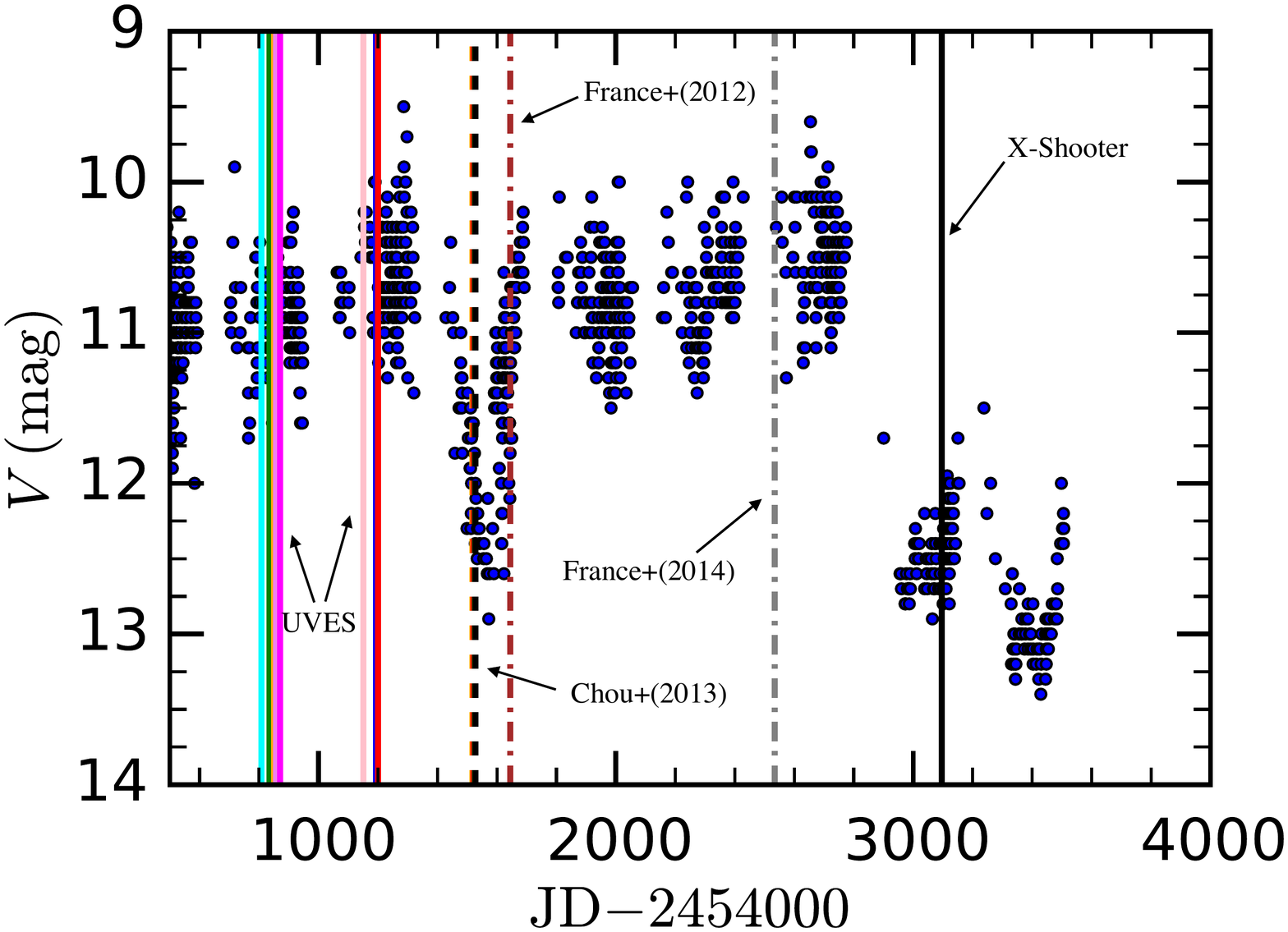}
\includegraphics[width=\columnwidth]{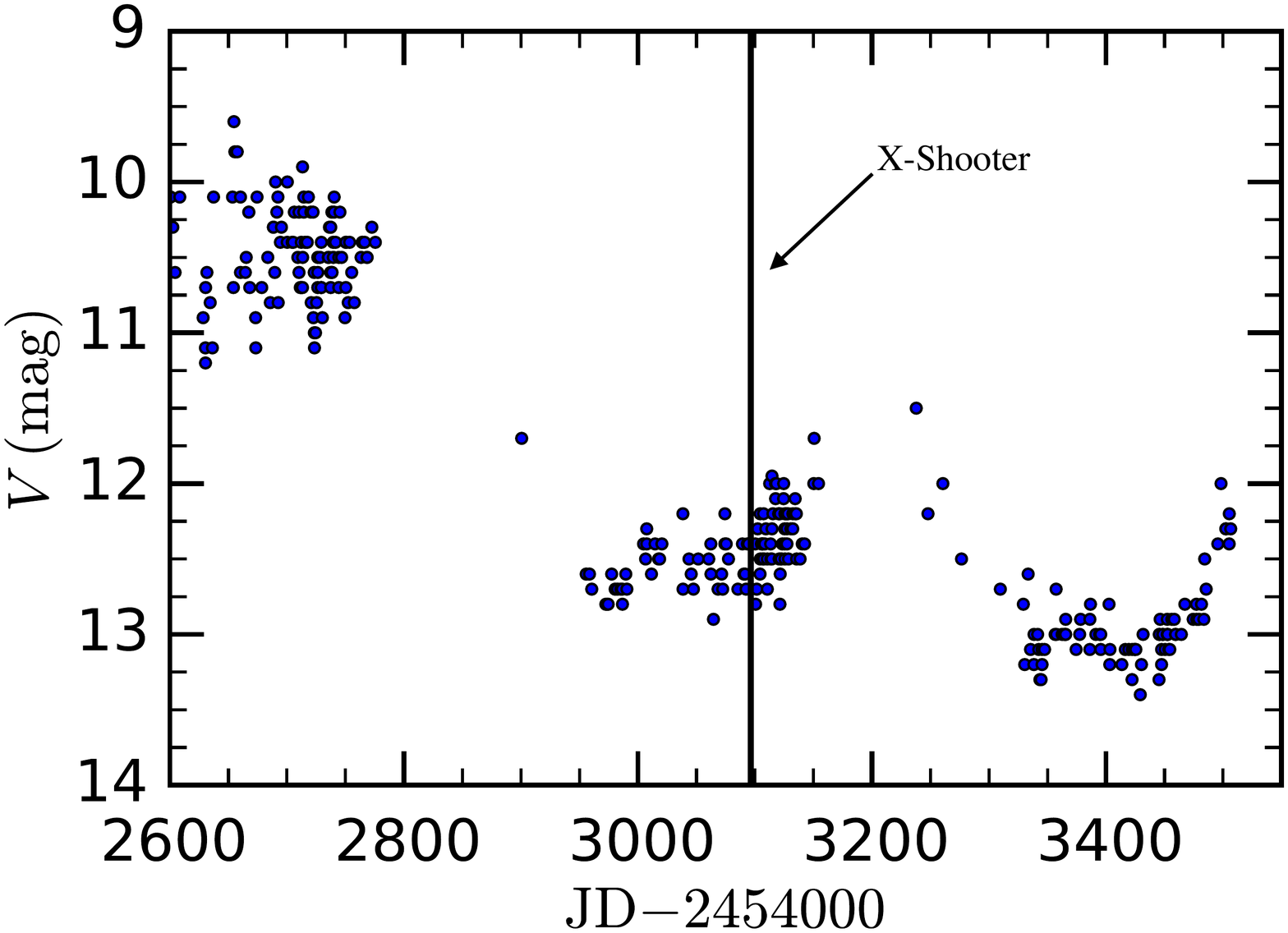}
\end{center}
\caption{Left panel: light curve in $V$-band of the unresolved RW Aur system from the AAVSO database. The solid vertical lines show the dates when the UVES and X-Shooter (black) spectra were taken (colour legend for UVES spectra is in Table \ref{tab:uves}). The dashed vertical lines show the dates of the four optical spectra by \citet{2013AJ....145..108C}, which overlap on the used scale. The brown and grey dashed-dotted lines show the dates of the two UV HST spectra by \citet[][respectively]{2012ApJ...756..171F,2014ApJ...794..160F}. Right panel:  
same as in left panel, zoomed during the second dimming event. Data in both panels are from \citet{aavso}.}
\label{fig:light_curve}
\end{figure*}

The epochs of our RW Aur A observations are compared against the system light curve in Fig.~\ref{fig:light_curve}. Our X-shooter data are taken well into the second dimming event. Since RW Aur A has been spectrally monitored at different wavelengths in the
past decade, in particular in the optical \citep{2005A&A...440..595A,2013AJ....145..108C,2015A&A...577A..73P,2016ApJ...820..139T} and in the UV \citep{2012ApJ...756..171F,2013ApJ...766...12M,2014ApJ...794..160F}, we can compare our observations with different datasets, obtained in both the bright and the dim states. In particular, we consider the following observables:

\begin{enumerate}
\item photometry and light curve of RW Aur A and B, to analyse the extinction properties of the occulting material;
\item atomic absorption lines against the stellar emission, to try to constrain the gas column density in front of the star, and obtain kinematic information on the gas lying along the line of sight;
\item photospheric absorption lines, to check whether the veiling has changed during the dimming;
\item accretion signatures in emission lines, to verify whether accretion properties have evolved significantly across the dimming events.
\end{enumerate}

\subsection{Photometric evolution of RW Aur}

RW Aur optical ($UBVRI$-bands) photometry has been monitored intensively for many decades now \citep[e.g.][]{2001A&A...375..977P,2013AJ....146..112R,2015A&A...577A..73P,2016AJ....151...29R}. However, very few of these photometrical measurements have resolved the two binary components. Since the primary star is significantly more massive, the optical photometry is usually dominated by it, and is thus sufficient to determine the occurrence of the dimming events. However, during the dimming events, star A can be dimmed by $>2$ magnitudes, and the unresolved optical photometry can be dominated by star B \citep[as has been noted by][]{2015A&A...584L...9S}. Since the X-Shooter spectra are flux calibrated and have only star A in the aperture, we can compute the synthetic optical photometry of the primary star on the date of the X-Shooter observations, thus checking how the colours and the magnitudes have evolved for star A from the bright to the faint state. The magnitudes of the $UBVRI$-bands are reported in Table \ref{tab:photometry} in the Johnson system \citep{1953ApJ...117..313J}.

In the literature, only two resolved photometric datasets of stars A and B are reported \citep{2001ApJ...556..265W,2015IBVS.6126....1A}. The first one was taken when RW Aur A had not shown any dimming event yet, whereas the second was taken during the second dimming event. We note that \citet{2015A&A...584L...9S} have also obtained resolved NIR photometry ($JHK$ bands) during the second dimming event.

\begin{figure*}
\begin{center}
\includegraphics[width=0.99\columnwidth]{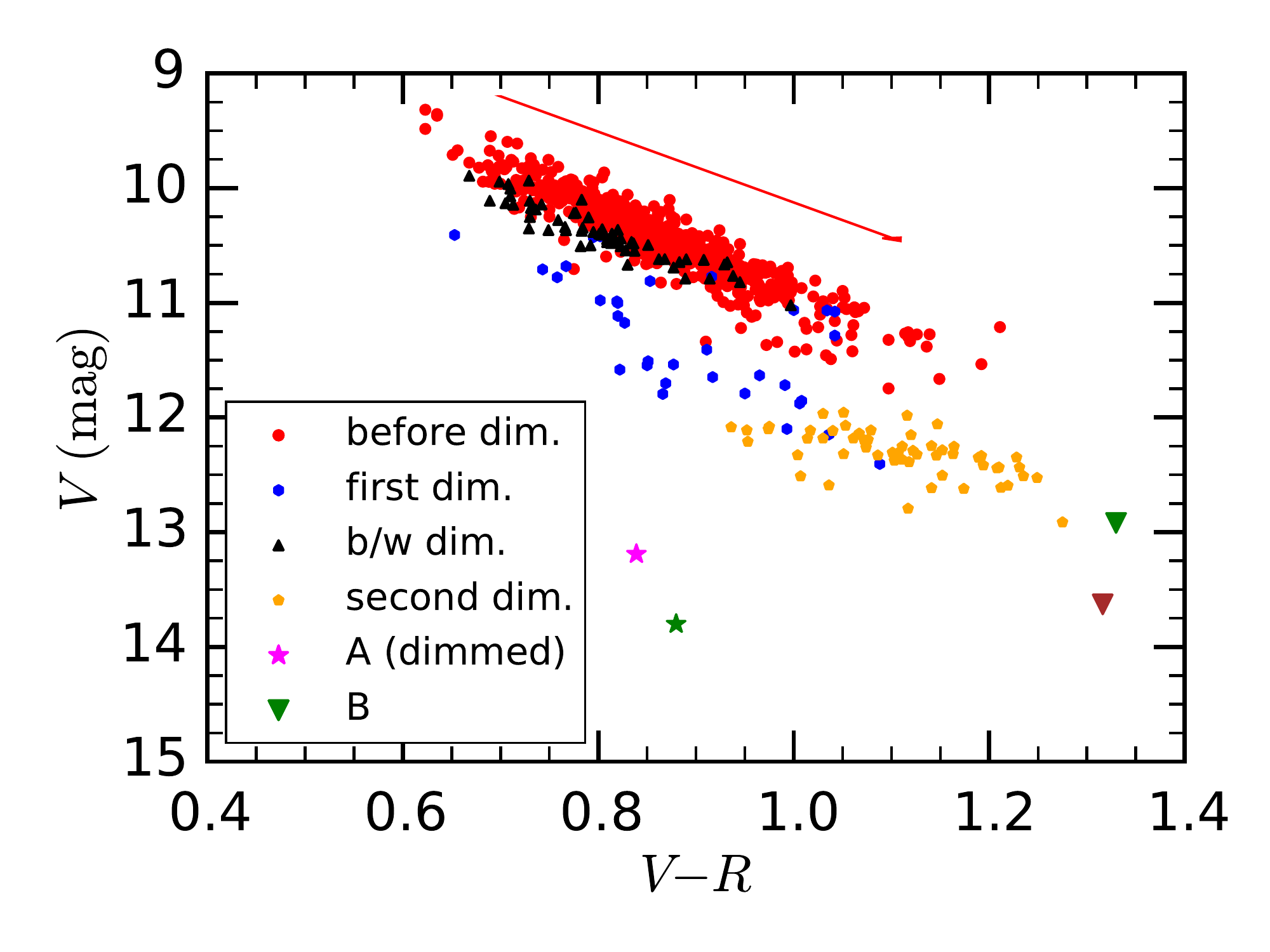}
\includegraphics[width=0.99\columnwidth]{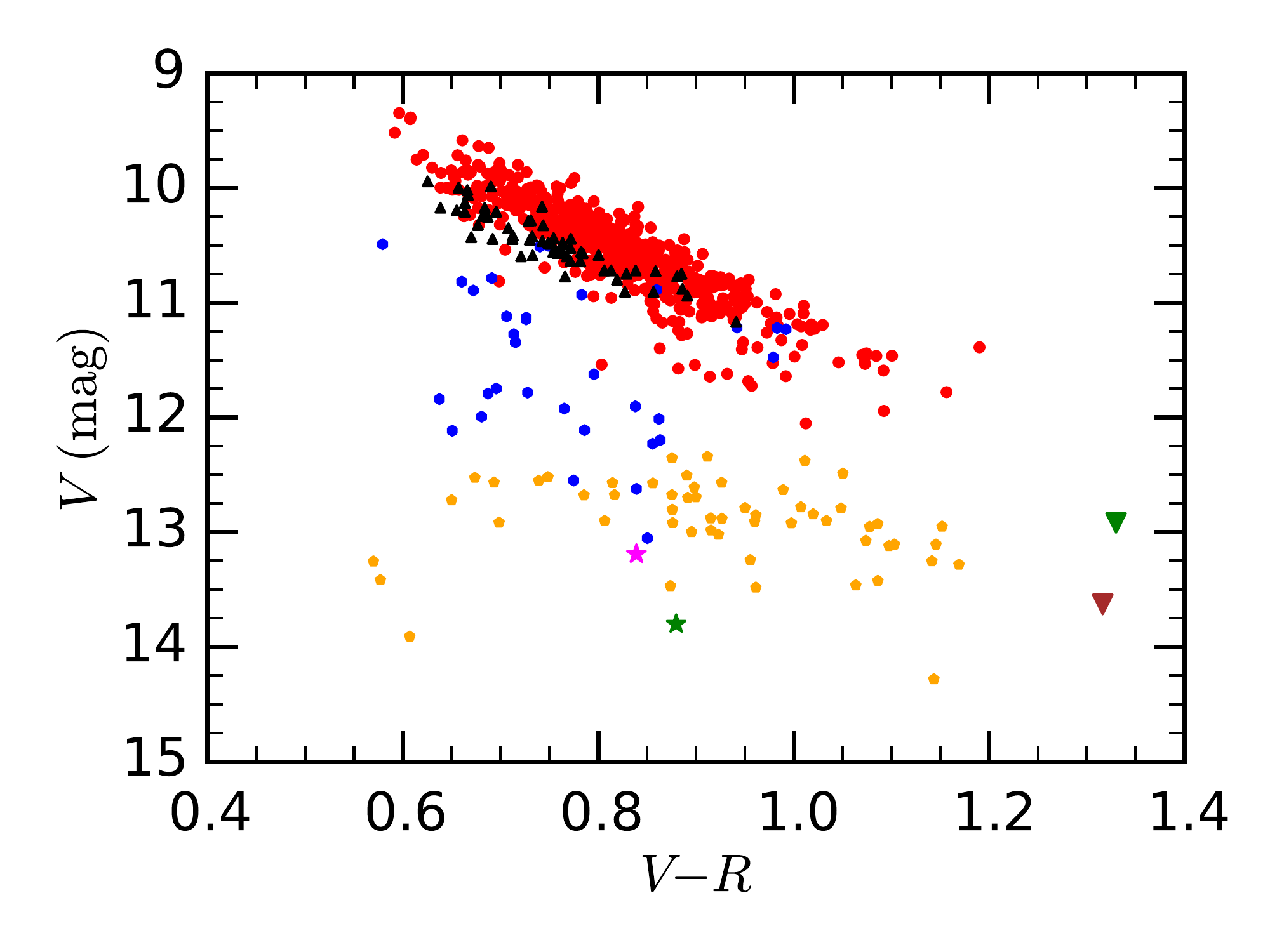}
\end{center}
\caption{Left panel: photometry of the unresolved RW Aur system A \citep[data from][]{2007A&A...461..183G,2015A&A...577A..73P}; the red arrow shows the reddening vector for typical ISM-like dust, for $A_V=1$ \citep{1977ApJ...217..425M}. Right panel: photometry of star A after subtraction of the averaged flux of star B \citep[taken from][]{2001ApJ...556..265W,2015IBVS.6126....1A} from the unresolved photometry. Legend: red dots are historical data from 1986 to 2005. Blue hexagons: first dimming. Black triangles: between the two dimming events. Orange pentagons: second dimming. Green star and green triangle: photometry of stars A and B, respectively, from \citetalias{2015IBVS.6126....1A}. Brown triangle: star B from \citetalias{2001ApJ...556..265W}. Magenta star: star A from our X-Shooter spectrum. All points are in the Johnson system \citep{1953ApJ...117..313J}.
}
\label{fig:photometry_1}
\end{figure*}

We then considered archival photometric data from \citet{2007A&A...461..183G} (from 1986 to 2005) and from \citet{2015A&A...577A..73P} (from November 8, 2011), where the system is unresolved. To reconstruct the photometric evolution of star A, we considered all the photometric points of the unresolved system, and we subtracted the flux of star B from them, assuming that it is stable with time. In Fig. \ref{fig:photometry_1} we show the photometry of the whole system (left panel), and the photometry of star A after subtraction of the flux of star B, which was calculated as the average of the estimates by \citet{2001ApJ...556..265W} and \citet[][right panel]{2015IBVS.6126....1A}. When star A is bright, the photometry does not change substantially, since star A dominates the flux of the system in the optical bands. The photometry of the two dimming events moves substantially in the V, V-R colour-magnitude diagram when we subtract the contribution from star B, and they become statistically similar to the two resolved photometric points of star A taken during the dimming. The large scatter in the corrected photometry traces the uncertainty on the unresolved photometry.

\begin{table}
\centering
\begin{tabular}{llccccc}
\hline
Star & Reference &  $U$  & $B$ &  $V$ & $R$ & $I$  \\
\hline
\hline
A & This paper & $14.00$ & $13.96$ & $13.19$ & $12.35$ & $11.55$ \\
A & \citetalias{2015IBVS.6126....1A} & $14.26$ & $14.50$ & $13.80$ & $12.92$ & $12.10$ \\
B & \citetalias{2015IBVS.6126....1A} & $14.97$ & $14.26$ & $12.92$ & $11.59$ & $10.49$ \\
\hline
\label{tab:photometry}
\end{tabular}
\caption{Resolved $UBVRI$ photometry of RW Aur A and B during the second dimming event from our X-Shooter spectrum and \citetalias{2015IBVS.6126....1A}. Values are in the Johnson system. The errors on our photometry are evaluated to be at the 2\% level.}
\end{table}

During the dimming events, while being about $3$\,mag fainter, the primary star retains the same colour range, that is, the extinction law is roughly grey in the optical bands, as suggested by \citet{2015IBVS.6126....1A} and confirmed by \citet{2015A&A...584L...9S} in $JHK$ bands. We thus confirm this result with the resolved photometry from our X-Shooter spectrum and by subtracting the flux of star B from unresolved archival photometry. Moreover, we observe that the extinction is likely also grey {\it \textup{during}} the dimming events, while it is compatible with interstellar
medium (ISM) extinction when the star is bright, with $A_V\sim1$ \citep{2001A&A...375..977P}.

\subsection{Atomic absorption lines against the stellar emission}

\begin{figure}
\center
\includegraphics[width=\columnwidth]{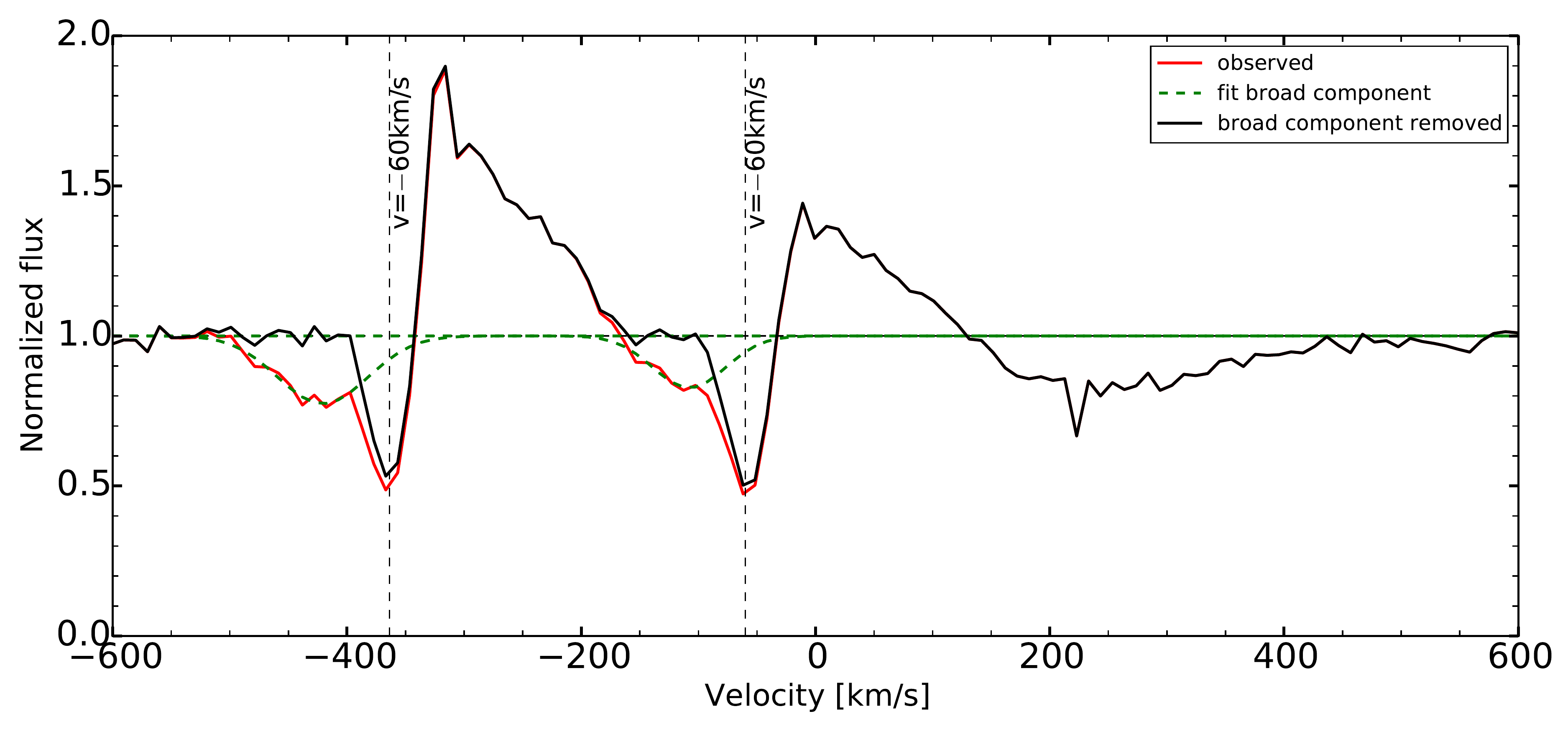}\\
\includegraphics[width=\columnwidth]{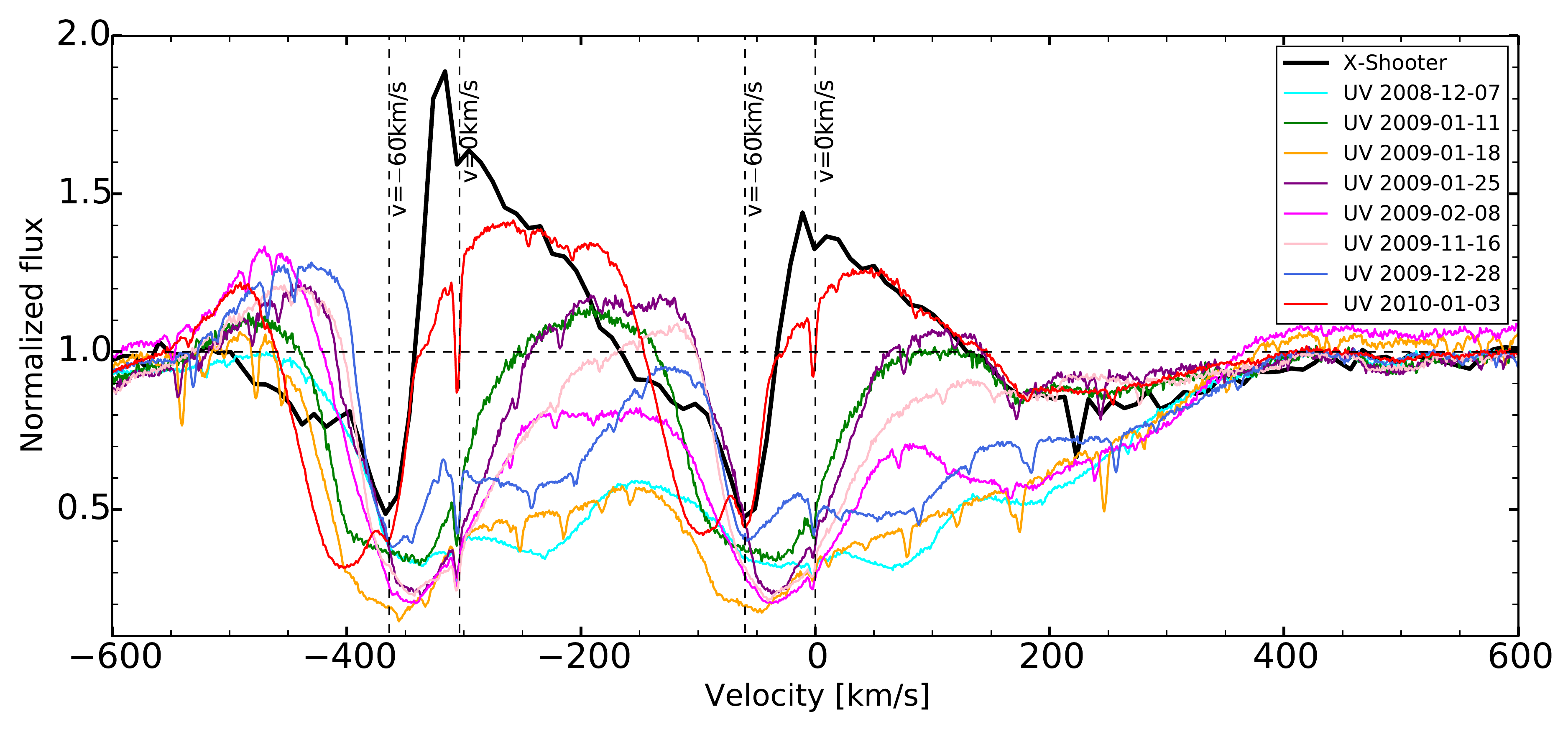}\\
\includegraphics[width=\columnwidth]{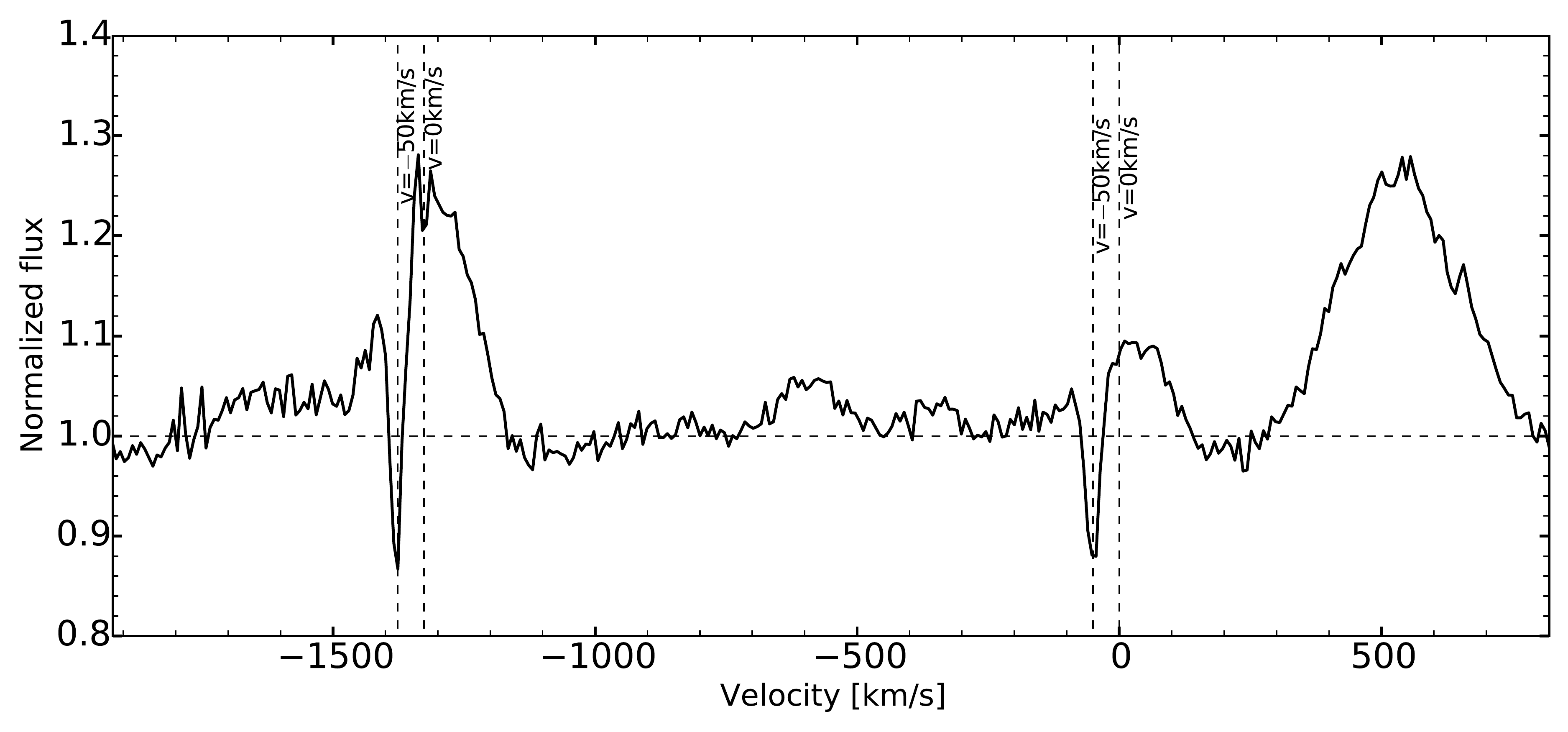}\\
\caption{Top panel: Na\,\textsc{i}\,D lines from X-Shooter. Central panel: Na\,\textsc{i}\,D lines from X-Shooter and UVES before the dimming events. Bottom panel: K\,\textsc{i}\,D lines from X-Shooter. All the velocities are heliocentric.}
\label{fig:ew}
\end{figure}

In the X-Shooter spectrum we clearly detect a narrow absorption component in the resonant Na\,\textsc{i}\,D (at $\lambda\sim590\,$nm) and K\,\textsc{i}\,D (at $\lambda\sim765$ and $770\,$nm) lines (see Fig. \ref{fig:ew}). This narrow absorption component probes gaseous material along the line of sight. 

\begin{figure}
\begin{center}
\includegraphics[width=\columnwidth]{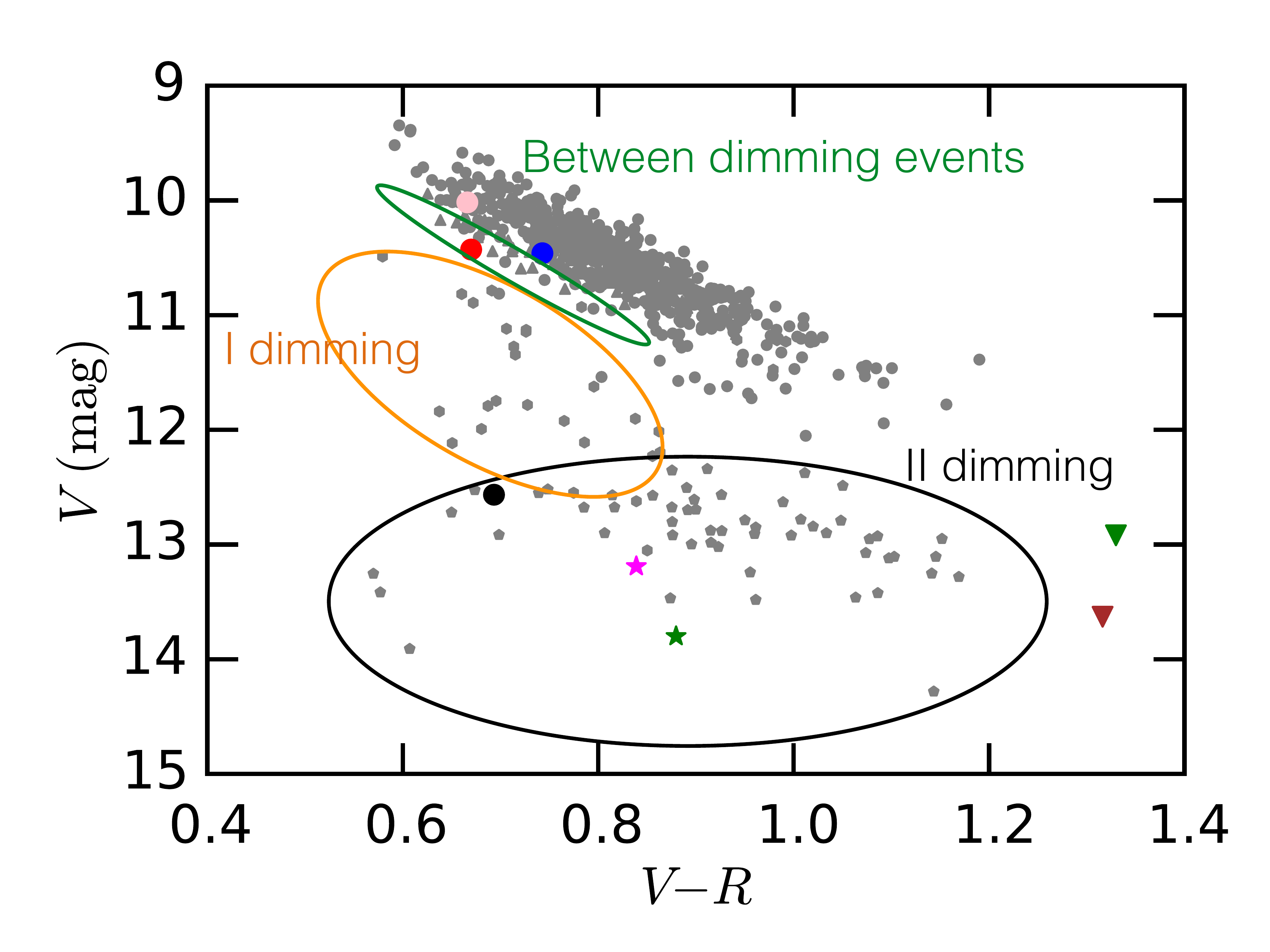}
\end{center}
\caption{Photometry of star A, as in the right panel of Fig. \ref{fig:photometry_1}. 
Legend: grey dots are historical data from 1986 to 2005. Grey hexagons: first dimming. Grey triangles: between the two dimming events. Grey pentagons: second dimming. Green star and green triangle: photometry of stars A and B, respectively, from \citetalias{2015IBVS.6126....1A}. Brown triangle: star B from \citetalias{2001ApJ...556..265W}. Magenta star: star A from our X-Shooter spectrum. All points are in the Johnson system. The coloured filled circles show photometric points closer than two days to the spectrographic observations with UVES and X-Shooter. The colour code is reported in Table \ref{tab:uves}.
}
\label{fig:photometry}
\end{figure}

\subsubsection{Equivalent widths and column densities}

For the Na doublet, we estimated the equivalent width (EW) of the narrow absorption feature assuming a Gaussian absorption for the broad hot component (Fig.~\ref{fig:ew}, top panel). We obtain an EW of $\sim0.3-0.6$\,\AA\,for the two Na narrow lines (depending on the modelling of the emission line), and of $\sim0.2$\,\AA\,for the two K lines. \citet{2015A&A...577A..73P} searched for narrow absorption signatures in the same Na lines, but they did not detect them, either because of a too low spectral resolution (they are      sensitive to EW$>0.2$\,\AA), or because of the intrinsic time variability of the physical phenomenon.

\begin{figure*}
\begin{center}
\includegraphics[width=0.66\columnwidth]{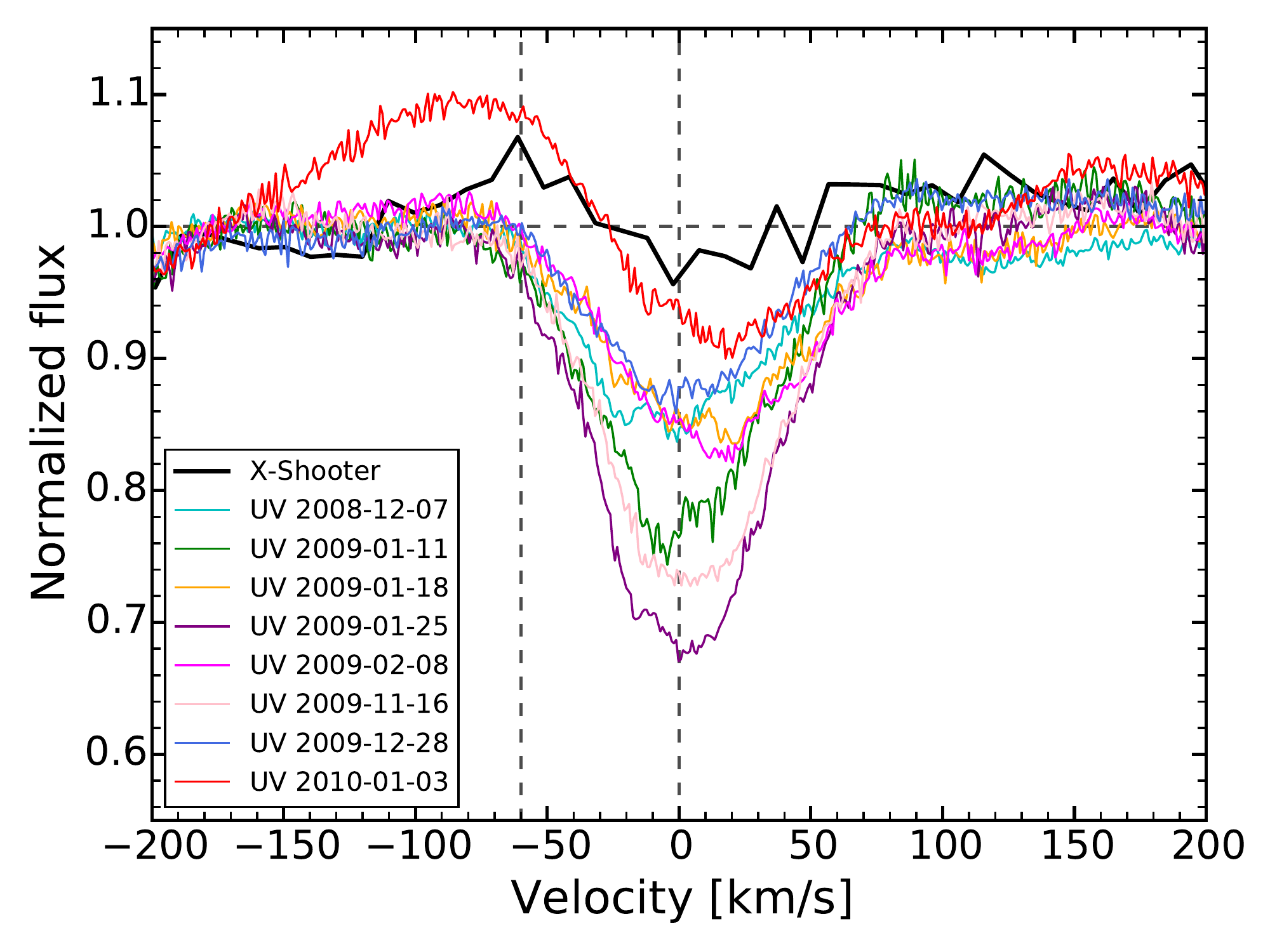}
\includegraphics[width=0.66\columnwidth]{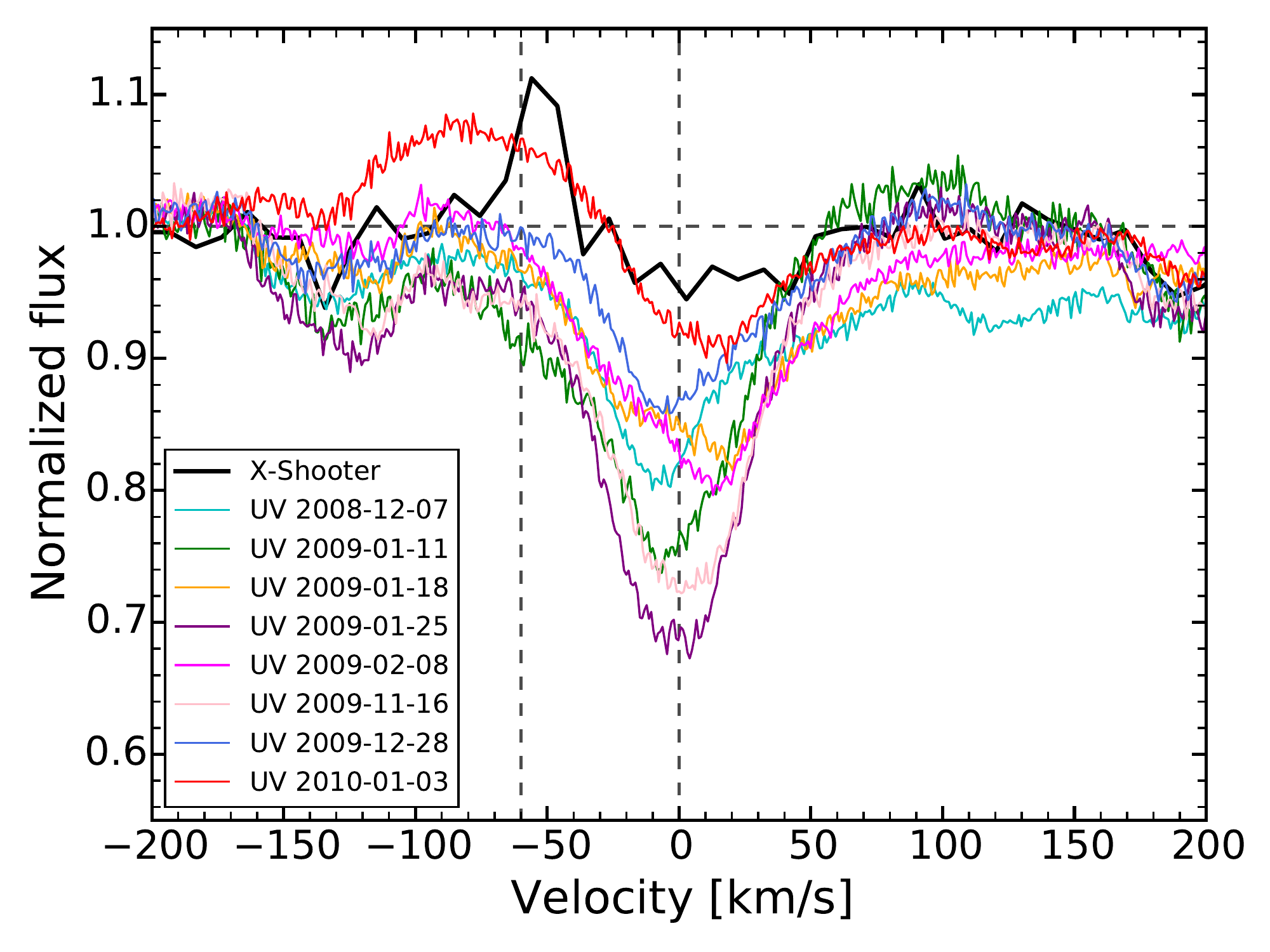}
\includegraphics[width=0.66\columnwidth]{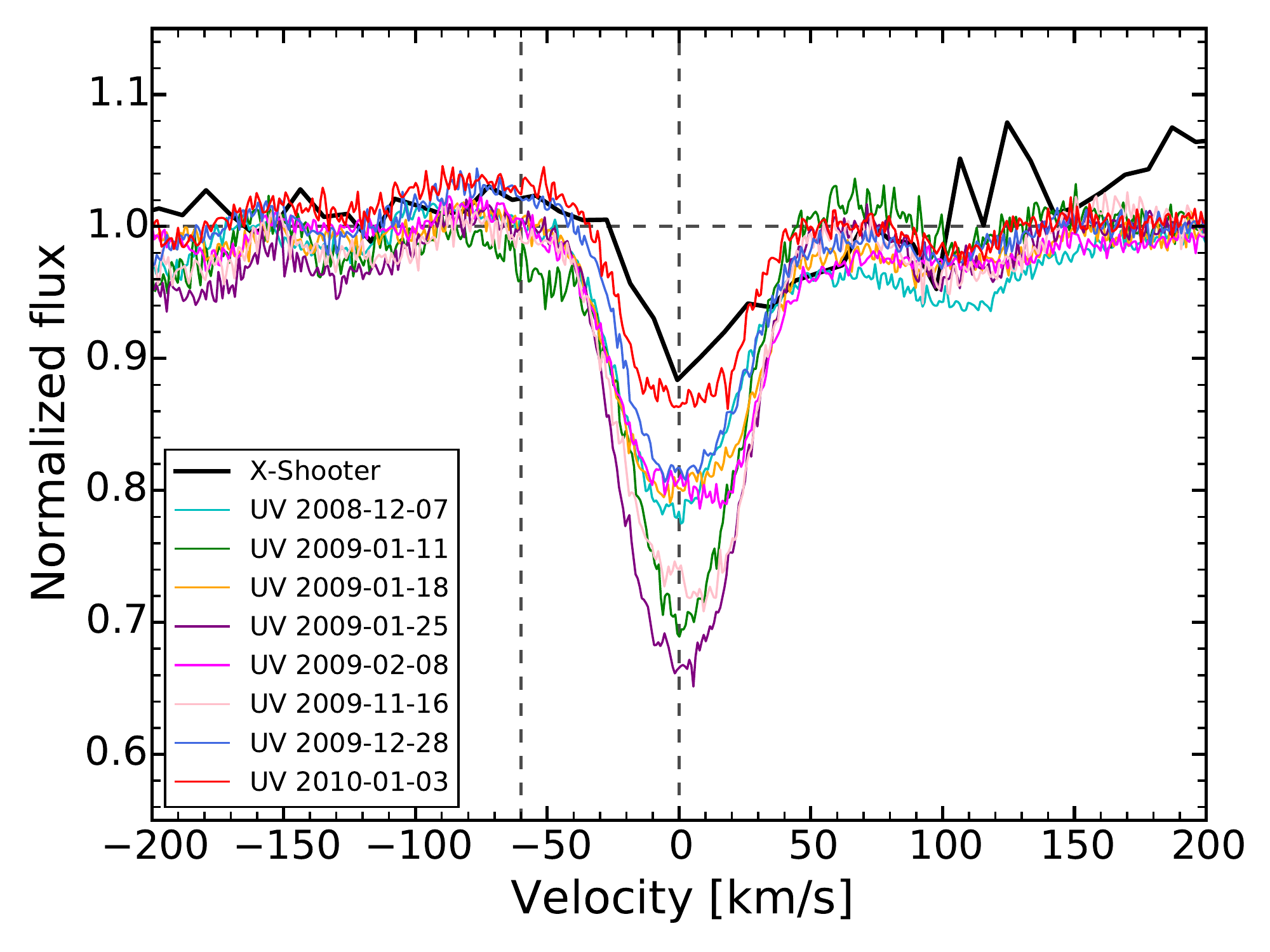}
\end{center}
\caption{Photospheric absorption lines of Ca\,\textsc{i} at 610.27 and 612.22\,nm (left and central panel, respectively) and Li at 670.78\,nm (right panel).}
\label{fig:photospheric_lines}
\end{figure*}

The fact that we obtain the same EW for both Na\,\textsc{i}\,D lines (see Fig. \ref{fig:ew}) indicates that the lines are saturated. The absorption oscillation strength $f$ of the two hyperfine transitions differs by a factor of $2$ \citep[e.g.][]{1973ApJS...26..333M}. By converting the EW of the Na\,\textsc{i}\,D$_1$ line at $589.75$\,nm (where the absorption oscillation strength is half the strength of the Na\,\textsc{i}\,D$_2$ line) into column density, we obtain a lower limit on the column density of atomic neutral sodium of $\sim3-6\times10^{12}$\,cm$^{-2}$. This value is a typical column density where the line saturates in ISM measurements \citep[e.g.][]{1975ApJ...202..634C}. The same procedure can be applied to the K\,\textsc{i}\,D$_1$ line, where we obtain $\sim10^{12}$\,cm$^{-2}$ as a lower limit on the column density of atomic neutral potassium. For the K doublet, the moderate spectral resolution cannot discern whether the lines are saturated. We also searched for absorption lines at the sodium and potassium ultraviolet doublets at $3300$ and $4045$\,\AA. These doublets are often used to determine the column density in the ISM when the Na\,\textsc{i}\,D lines are saturated \citep[e.g.][]{1975ApJ...202..634C}, since the absorption oscillation strengths are lower by a factor of $\sim10-100$ \citep[e.g.][]{1973ApJS...26..333M}. However, we did not detect these absorption lines. As a result of the strong accretion signatures and the poorer spectral resolution in the UVB arm, we have a detection limit of absorption lines with EW$\lesssim0.27$\,\AA. This yields an upper limit on the column density of neutral atomic sodium of $\sim2\times10^{14}$\,cm$^{-2}$.

\subsubsection{Line profiles}

\label{subsubsec:profiles}

\citet{1999A&A...352L..95G} estimated the radial velocity (RV) of star A as $v_0=15.87\pm0.55$\,km/s. Recent studies are in agreement with this estimate \citep{2001A&A...365...90F,2016ApJ...820..139T}. We note that the heliocentric velocity oscillates by $\pm5-6$\,km/s \citep{1999A&A...352L..95G,2001A&A...369..993P,2016ApJ...820..139T}, and \citet{2001A&A...369..993P} noted that these variations are periodic (with a period of $2.77$ days). We took this value of RV into account to derive the velocity relative to the star of the Na and K lines in our spectra. We obtain that the former is $-60\pm10$\,km/s, while the latter is $-51\pm10$\,km/s (see Fig. \ref{fig:ew}), where the uncertainties are dominated by the S/N. These absorption lines are thus probing moderately blueshifted material. Both lines are likely tracing the same material, since they have a compatible velocity.

We then studied the Na\,\textsc{i}\,D line profiles of the eight high-resolution UVES spectra of RW Aur A (see central panel of Fig. \ref{fig:ew}). They are broader and significantly different from the line profile obtained with X-Shooter. In particular, there is much less redshifted absorption (from $0$ to $+200$\,km/s) in the X-Shooter data than in (most of) the UVES data, which suggests either that the mass accretion rate has decreased in the X-Shooter data or that the accretion region is now partly occulted. Another main difference is the high-velocity blueshifted absorption (from about $-100$\,km/s to about $-200$\,km/s) seen in the X-Shooter spectrum that is not seen earlier, which suggests the development of a fast wind. The 2010 UVES spectrum is the most similar to the X-Shooter one, with a typical P Cyg profile. Only this UVES spectrum shows a clear absorption component at $\sim-60$\,km/s (see Fig. \ref{fig:ew}, central panel). During this epoch, the Na\,\textsc{i}\,D doublet appears not to be saturated. We note, however, that in the colour-magnitude diagram (Fig.~\ref{fig:photometry}) this spectrum seems to be correlated with a minor dimming event. A similar feature was also observed in some older spectra, such as those shown in Fig. 2 of \citet{2005A&A...440..595A}. The same velocities ($\sim-60$\,km/s) in absorption have also been observed in Ca\,\textsc{ii} lines at the beginning of the first dimming event in 2010 by \citet{2013AJ....145..108C}. Interestingly, \citet{2013AJ....145..108C} obtained four different spectra during the ingress time of the first dimming event. It is apparent that there is a very fast transition from a line profile similar to those in the UVES spectra presented here, and a line profile that closely resembles our X-Shooter profile. Similar moderately blueshifted lines at $\sim-60$\,km/s have also been observed by \citet{2015A&A...577A..73P} in the K optical doublet and by \citet{2014ApJ...794..160F} in low-ionisation UV lines. All these observations suggest that during the bright state the lines are dominated by accretion properties, whereas during the dim states they are dominated by wind features.

\subsubsection{Narrow co-moving component}

In all the UVES spectra of the Na\,\textsc{i}\,D lines we observe a much narrower unsaturated absorption component (barely detected by X-Shooter) at $\sim-2$\,km/s relatively to star A (see central panel of Fig. \ref{fig:ew}). This absorbing material is consistent with moving coherently with star A (within the uncertainty on the estimate of the velocity of the star). These lines are unlikely to be photospheric, since they are very narrow. They present an EW of $\sim0.017\,$\AA\, and $\sim0.030\,$\AA\,in the red and blue component, respectively, indicating a neutral sodium column density of $\sim1.7\times10^{11}$\,cm$^{-2}$. These velocities and equivalent widths are in agreement with the Na absorption lines caused by the atomic cloud surrounding the Taurus-Aurigae star-forming region \citep{2015ApJ...814...14P}.

\subsection{Veiling (?) of photospheric absorption lines}
\label{subsec:veiling}

We have three strong photospheric lines in the X-Shooter and in the UVES spectra: two Ca\,\textsc{i} lines (at $610.27$ and $612.22$\,nm), and the Li line at $670.78$\,nm (see Fig.~\ref{fig:photospheric_lines}). By comparing the UVES and X-Shooter lines, we clearly see that the EW of these photospheric lines during the dimming event decreases
strongly. This confirms the result by \citet{2016ApJ...820..139T}, who have observed the same trend on a large number of epochs. The 2010 (red) UVES spectrum again resembles the spectrum of X-Shooter, suggesting that whatever is causing the spectral changes related to the dimming event might be an extreme state of a more frequent phenomenon.

\subsection{Emission lines and accretion luminosity}

The high accretion rate of RW Aur A causes a plethora of strong emission lines related to the accretion process. Many of these are within the wavelength range covered by the X-Shooter spectrum \citep[e.g.][]{2014A&A...561A...2A}. In particular, we consider in the following the H$\alpha$, H$\beta$, H$\gamma$, H$\delta$, Pa$\beta$, Pa$\gamma$, Pa$\delta$, Br$\gamma$, Ca\,\textsc{ii}\,K, the Ca\,\textsc{ii} IR triplet, and He\,\textsc{i} at $587$ and $667$\,nm lines to derive an estimate of the accretion rate during the dimming.

\subsubsection{Accretion luminosity}

\begin{figure}
\begin{center}
\includegraphics[width=\columnwidth]{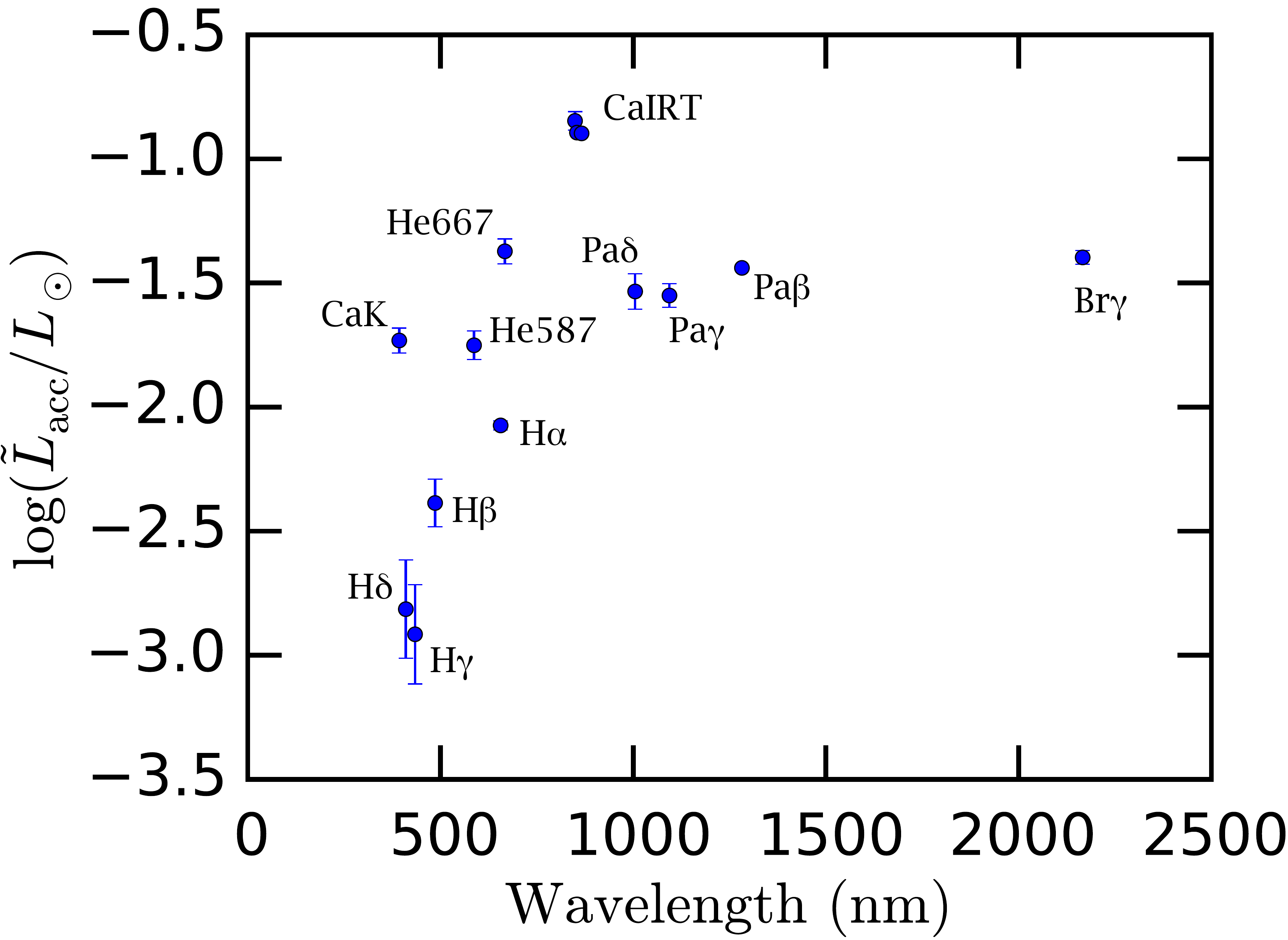}
\end{center}
\caption{Apparent accretion luminosities (accretion luminosities not corrected for extinction) from various emission lines labelled in the plot. The relations between $\tilde{L}_{\rm acc}$ and $L_{\rm line}$ are taken from \citet{2014A&A...561A...2A}.
}
\label{fig:acc}
\end{figure}

The X-Shooter spectrum we obtained was flux calibrated, and the observed flux of the emission lines was then obtained by direct integration of the continuum-subtracted spectrum across the line. This was converted into observed line luminosity ($\tilde{L}_{\rm line}$) using the distance of 140 pc to RW Aur A. Using the relations between line luminosity and accretion luminosity by \citet{2014A&A...561A...2A}, we obtain the apparent accretion luminosity $\tilde{L}_{\rm acc}$. This is shown for each line in Fig. \ref{fig:acc}, where the lines are ordered according to their central wavelengths. We refer to this quantity as apparent luminosity since we have not corrected for extinction, as we discuss in the following. The apparent accretion luminosity as derived from the Balmer lines is much lower than the luminosity obtained from the Paschen lines. Although the sign of this variation is consistent with the sign generated by dust extinction, this interpretation is inconsistent with the fact that the photometric colours indicate grey extinction up to the $K$ band \citep{2015IBVS.6126....1A,2015A&A...584L...9S}. Moreover, the large scatter of $\sim1$\,dex between $\tilde{L}_{\rm acc}$ measured from lines located close in wavelengths, such as Ca\,\textsc{ii}\,K and H$\beta$, is too large to be simply ascribed to differential extinction. Instead, Fig.~\ref{fig:acc} is most readily explicable in terms of {\it gas} absorption, which would predict stronger attenuation of the Balmer series than the Paschen lines.

We  now consider a grey extinction of $A_V=3$\,mag \citep{2015A&A...584L...9S} to recover the actual accretion luminosity $L_{\rm acc}$. We use the following stellar parameters: $M_*=1.4M_\odot$, $R_*=1.6R_\odot$ , and $\log{T_{\rm eff}}=3.67$ \citep[][and references therein]{2013AJ....146..112R}. If we do not take into account the Balmer lines and average across all the other lines, we obtain a mass accretion rate $\dot{M}\approx4\times10^{-8}M_\odot/$yr, in agreement with the estimate of $2\times10^{-8}M_\odot/$yr by \citet[][since accretion rates are considered to be accurate to within a factor of two, and these authors used a different method to estimate the accretion rate, namely NUV and optical excess and a template stellar spectrum]{2013ApJ...767..112I}. The accretion rate is much lower than the rate obtained by \citet{1995ApJ...452..736H} ($\sim10^{-6}M_\odot/$yr), but we note that they used different stellar parameters (a larger and less massive star), which would yield an accretion rate of $\sim2\times10^{-7}M_\odot/$yr from our accretion luminosity. However, we cannot necessarily conclude that the accretion rate has decreased since the 1990s, since the lines used in our estimate (i.e. all those in Fig.~\ref{fig:acc} except for the Balmer lines) might also be affected by gas absorption.

\subsubsection{He\,\textsc{i} emission line at $587$\,nm}

\begin{figure}
\center
\includegraphics[width=\columnwidth]{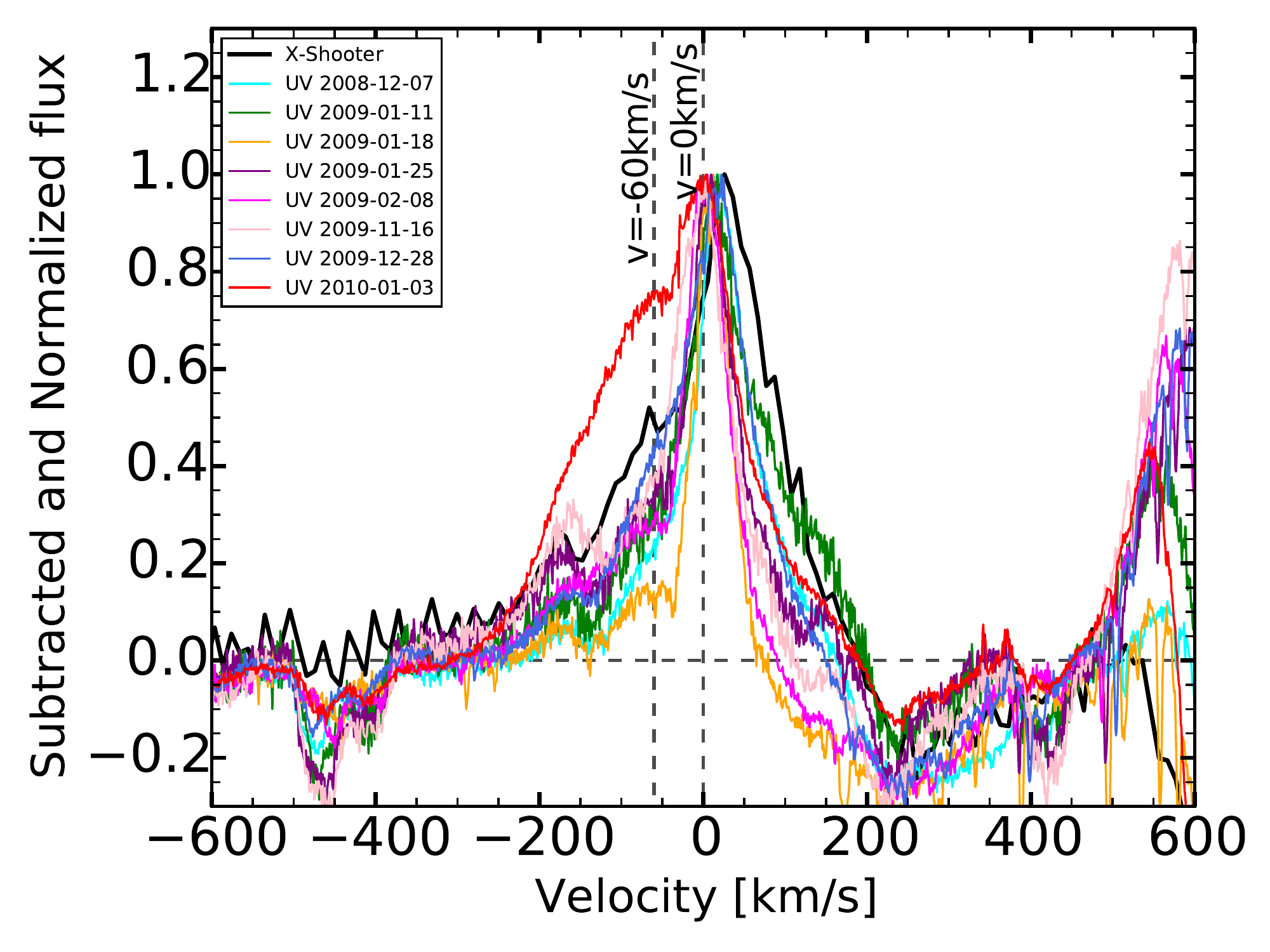}\\
\caption{Continuum-subtracted and normalised emission lines of He\,\textsc{i} at 587.56\,nm in the heliocentric reference frame.}
\label{fig:emission_lines}
\end{figure}

We analyse the profile of the He\,\textsc{i} line at $587$\,nm
in more detail. This line is also present in the UVES spectra. The continuum-subtracted peak-normalised profiles of the line are shown in Fig. \ref{fig:emission_lines}. The line width is relatively steady across the dimming event. This is also observed in the H$\alpha$ emission line. The only exception is the 2010 UVES case (and marginally in the X-Shooter one), which shows enhanced emission at $-60$\,km/s in the heliocentric reference frame. We note that an emission feature at the same velocities is also observed in the Ca\,\textsc{i} lines shown in Fig.~\ref{fig:photospheric_lines}. The relatively steady width of this line suggests that the accretion rate has not significantly increased during the dimming event. In particular, the variations seen in the broad component of the 2010 UVES and in the X-Shooter spectra are likely due to an enhanced hot wind \citep{2001ApJ...551.1037B,2011MNRAS.411.2383K}, whereas the narrow component tracing accretion does not seem to be affected by the dimming.

\section{Discussion}
\label{sec:discussion}

The peculiar photometric and spectroscopic behaviour of RW Aur A has led to different interpretations of its variable nature, and in particular of its dimming events. Here we analyse the possibilities presented in the literature, and compare how the spectra presented here either confirm or discard these hypotheses. We finally propose a new interpretation.

\subsection{Tidally disrupted material}

Soon after the discovery of the first dimming event, \citet{2013AJ....146..112R} interpreted the drop in optical flux as due to the outer disk being partially disrupted by the outer binary. \citet{2015MNRAS.449.1996D} analysed the primary disk properties and the proper motions of RW Aur A and B in detail, concluding that the system is well represented by a parabolic tidal encounter between the primary disk and star B. Furthermore, they showed that with such dynamics, part of the disrupted disk would form a bridge structure between the two stars, and that it would precess across the line of sight to star A, thus leading to dimming events. Finally, \citet{2016AJ....151...29R} interpreted the observed colour evolution during the ingress time of the dimming events as a sign of dust segregation within the occulting body.

However, two observations suggest that this theory does not explain the recent dimmings. First, as mentioned in Sect. \ref{sec:intro}, the optical and NIR dimmings, observed to be grey up to  $\sim K$ band, show a strong correlation with IR excess emission in $L$ and $M$ bands \rev{\citep{2015IBVS.6143....1S,2016MNRAS.463.4459B}}, compatible with black-body emission at $\sim1000$\,K, thus with dust thermal emission at $\sim0.1$\,AU from the central star. This IR excess indicates that the dimming is correlated to some major perturbation of the inner disk. Moreover, the hydrodynamical simulations by \citet{2015MNRAS.449.1996D} indicate that the material predicted to lie close to the line of sight to star A has a heliocentric RV between $-8$ and $0$\,km/s. The increase in dust column density would be correlated to an increase in gas column density, since the dynamics of the encounter is mainly ballistic \citep{2015MNRAS.449.1996D}. However, in our X-Shooter spectrum we do not observe any increase in the gas column density at such low velocities (see the Na and K absorption lines in Fig. \ref{fig:ew}).

From the IR excess, we can conclude that the occulting material is very close to the central star \citep{2015IBVS.6143....1S}. Moreover, the lower apparent accretion luminosities derived for the Balmer lines compared to other lines tracing accretion indicate that gaseous material is obscuring the accretion regions. \citet{2015A&A...577A..73P,2016ApJ...820..139T} suggested that [O\,\textsc{i}] and [S\,\textsc{ii}] forbidden lines do not seem to change substantially in total flux between the bright and dim state, indicating that winds and jets on larger scales are not significantly affected by the obscuration, therefore limiting the spatial extent of the occulting material.

\subsection{Dust-laden stellar wind}

\citet{2015A&A...577A..73P} and \citet{2015IBVS.6143....1S} interpreted the dimming events as caused by dust-laden stellar wind, that
is, by a strong stellar wind eroding the inner dusty disk, and dragging dust grains as large as $\sim1$\,$\mu$m along the line of sight. These large sizes are required to reproduce the grey extinction in the optical and NIR bands. In this scenario, the occultation would be a consequence of a stellar outburst; the brightening of the system at $L$ and $M$ bands could then plausibly be ascribed to an episode of enhanced accretion that may trigger an episode of enhanced outflow \citep[e.g.][]{2014ApJ...796...29S}. However, our observations do not indicate a significant accretion variation between the faint and bright periods. In particular, the He\,\textsc{i} and H$\alpha$ emission lines do not show significant variation in their profiles. \citet{2016ApJ...820..139T} suggested that new estimates of the mass flux within the jet might be a stronger indication of whether such a stellar outburst has indeed occurred. With the available observations, it seems difficult to either confirm or discard this hypothesis. 

\subsection{Warped and puffed up inner disk}

Another possibility, which we prefer based on available data, is that the enhancement of $L$ and $M$ band emission \citep{2015IBVS.6143....1S} is due not to an accretion burst, but to a reconfiguration of the inner disk\rev{,} associated with an enhancement of the {\it \textup{disk}} wind \rev{signatures \citep[see also][]{2016MNRAS.463.4459B}}. A puffing up of the inner disk would increase the solid angle of dusty material subtended at the star and so enhance the re-radiated infrared  flux without the need to invoke an episode of enhanced accretion. It is notable that the likely location of the material emitting in the $L$ and $M$ band is at $\sim 0.1$\,AU, where the dynamical timescale is of the order of a few days, comparable with the timescale for the onset of the optical dimming. If the disk is suitably aligned with respect to the observer's line of sight, this puffing up of the inner disk (possibly due to hydrodynamical instabilities triggered by the tidal encounter, or due to an interaction between the stellar wind and the inner disk) could also occult the star and X-ray emitting region, reducing the optical flux and accounting for the high gas column detected in the X-ray spectrum \citep{2015A&A...584L...9S}. At minimum light, the residual optical flux could be dominated by scattered photons from the vicinity of the stellar polar caps and thus yield a nearly continuous spectrum, consistent with the loss of photospheric absorption features noted by \citet{2016ApJ...820..139T} and in Sect. \ref{subsec:veiling} of this paper during the dimming events. \rev{The polar caps are expected to be characterised by very high effective temperature in the magnetospheric accretion scenario (with $T_{\rm eff}\gtrsim10000\,$K, much hotter than the rest of the photosphere), thus leading to more pronounced veiling. Another possibility is that during the photometric minima, the photospheric lines are veiled by the corresponding emission lines from the circumstellar material (e.g. the observed jet), as has been directly observed in the UX Ori type star RR Tau \citep{2002ApJ...564..405R}.} If the puffing up of the disk is associated with enhanced mass loading of the disk wind, then this also naturally explains the appearance of strong absorption features at $- 60\,$km/s. \rev{However, the lack of correlation between the appearance of the disk wind signatures and the accretion properties of the star may be indicating that the wind does not exhibit an actual enhanced mass-loss rate \citep[see also the discussion in][]{2016MNRAS.463.4459B}, but the puffed-up inner rim has simply displaced the base of the flow closer along the line of sight, thus yielding the blueshifted absorption for geometrical meanings.} We observe \rev{the $-60\,$km/s feature}  in absorption in the Na and K optical doublets (Fig. \ref{fig:ew}) and in emission in the He\,\textsc{i} line (Fig. \ref{fig:emission_lines}). The same wind velocities have been observed in the 2013 HST data by \citet[][when RW Aur was between the two dimming events]{2014ApJ...794..160F}, who clearly illustrated the presence of a low-ionisation wind between $-80$ and $-30\,$km/s in absorption lines of N\,\textsc{i}, Si\,\textsc{ii,} and Si\,\textsc{iii}. In this work, we have provided an upper and a lower limit for the column density of this wind in atomic sodium. It is notable that the observed velocities of absorption features are of the order of the escape velocity at $\sim 0.1\,$AU: these velocities exceed those of thermally driven disk winds and point instead to a magnetically driven wind for which the flow velocity is of the order of the local escape velocity.

This scenario requires that the disk is seen close to edge-on, a possibility discarded by \citet{2013AJ....146..112R} based on the inclination of the outer disk ($45 - 60^\circ$), as estimated from mm interferometric observations \citep{2006A&A...452..897C}. However, NIR interferometric observations by \citet{2007ApJ...669.1072E} found the inner disk to be more inclined ($77^\circ$). Moreover, \citet{2013ApJ...766...12M} estimated the inner disk inclination to be $>60^\circ$, based on the observation of CO absorption lines in their UV spectra. The inconsistency between these inclinations might come from a warped inner disk\rev{, as noted also by \citet{2016MNRAS.463.4459B}}. If this is the case, a puffed-up inner disk could lie along the line of sight and occult the primary star. Moreover, the puffed-up ring would intercept more stellar photons and re-radiate them at the local temperature, thus explaining the IR excess observed by \citet{2015IBVS.6143....1S}. A magnetically driven warp \citep[][]{2011MNRAS.412.2799F}, as in the case of AA Tau \citep[e.g.][]{2007A&A...463.1017B,2013ApJ...762...40C}, is unlikely to occur for a $1.4M_\odot$ star, since the magnetic dipole moment is expected to decrease steeply with mass in post-main-sequence
(PMS) stars \citep[e.g.][]{2012ApJ...755...97G}. A warped structure could be produced by a close spectroscopic binary misaligned to the outer disk \citep[e.g.][]{2013MNRAS.433.2142F,2014MNRAS.442.3700F}. Such a binary has been proposed by \citet{2001A&A...375..977P} to model a $2.77$ -day variability observed in some emission line profiles \citep[even though this hypothesis has been contested by other authors, e.g.][]{2012AstL...38..167D}. We note that warped inner disks caused by the secular interaction with a (sub-)
stellar companion are now observed, for example in the HD142527 system \citep{2015ApJ...798L..44M,2015ApJ...811...92C}. Moreover, other systems with a low inclination or almost face-on outer disks are now showing minor dimming events \citep{2016MNRAS.462L.101A,2016MNRAS.463.2265S}. However, the origin of these dimmings might be different from the dimming of RW Aur A.

This scenario of a puffed-up inner disk \rev{(and possibly enhanced wind mass-loss)} can qualitatively account for the optical dimming, enhanced gas column, and blueshifted absorption features in atomic lines, as well as for the loss of photospheric absorption features during dimming events. In this scenario the lack of significant colour variations then results from a coincidental combination of the effect of reddening on an intrinsically blue-scattered light spectrum.

As mentioned in Sect. \ref{subsubsec:profiles}, \citet{2013AJ....145..108C} serendipitously obtained four different spectra of the system during the ingress time of the 2010 dimming event (see Fig. \ref{fig:light_curve}), with 2-3 days separation from one spectrum to another. The first two spectra are similar to our UVES spectra, whereas the second two are very similar to the X-Shooter one. The fact that the dimming starts before the spectral variations suggests that the disk's geometrical changes give rise to \rev{either} an enhancement of the wind mass-loss rate, \rev{or simply a geometrical displacement of the base of the disk wind against the spectral continuum}. The lag in wind response is of the order of the local dynamical time, as expected for a magnetically driven disk wind.

If the occulting material in the inner disk only obscures the central star, then the optical signal observed during the dimming is mainly scattered by such material. As scattered light has an increased polarisation signature, this scenario could be confirmed by measuring an enhancement of the amount of polarisation with respect to observations taken before the first dimming \citep{2005MNRAS.359.1049V}. A similar analysis has been performed in different systems: an enhancement of the polarisation signal of up to 10\% is seen in the dim state of various higher mass stars, the UX Ori young variable stars \citep{1998AstL...24..802G}, and this is explained as an effect of sporadic obscuration of the star by material in its circumstellar disk \citep{2003ApJ...594L..47D}.

\subsection{Similarities to other systems}

Dimming events similar to those observed for RW~Aur A have recently
been reported for AA~Tau and V409~Tau \citep{2013A&A...557A..77B, 2015AJ....150...32R}. 
Both systems show dimming events with $\gtrsim 2\,$mag in V and durations $\gtrsim 100\,$days, which have been postulated to be caused by obscuration of the central star by disk features rotating into view. In particular, from an analysis of molecular disk emission lines in the AA~Tau system, \citet{2015A&A...584A..51S} showed that the disk emission within $\sim1\,$AU has varied across the dimming event, whereas little or no variation in the emission from the outer disk was observed in the outer disk. 
These models argue in the same direction as our proposed explanation for RW~Aur A, namely a disturbance in the inner disk. The cause for this disturbance  (e.g. a magnetic interaction or a close low-mass binary companion) might differ. Nevertheless, the analogy of the observed absorption signatures, rather grey extinction, and an increase in gas column density of the order of $10^{22}$\,cm$^{-2}$, suggests that their inner disks share many similarities.

\section{Conclusions}
\label{sec:conclusions}

We have analysed and compared X-Shooter and UVES spectra of RW Aur A across the two recent dimming events. Our study yields the following conclusions:

\begin{itemize}
\item We confirm previous analyses stating that the extinction of RW Aur A during the second dimming event is compatible with being grey.
\item The optical Na and K doublets during the dim states show clear signatures of a blueshifted inner disk wind at $\sim-60\,$km/s.
\item We confirm earlier studies showing that RW Aur A spectrum looks highly veiled during the dimming events.
\item No major accretion variation is observed across the dimming events.
\end{itemize}

To interpret these results together with many other observations of the system, we have proposed a new mechanism that might be at the origin of the dimming events. In particular, we propose that these dimming events trace major perturbations of a misaligned
or warped inner disk. This interpretation would collocate RW Aur A in a class of objects where the inner disk has caused significant photometric and spectrometric evolution of the hosted protoplanetary disk, such as the famous cases of AA~Tau and V409~Tau \citep{2013A&A...557A..77B, 2015AJ....150...32R}. Interestingly, RW Aur A has the additional peculiarity of having recently suffered a perturbing tidal encounter with its companion star RW Aur B.

\begin{acknowledgements}
\rev{We thank the anonymous referee, whose comment helped clarifying part of the paper}. We \rev{are grateful to} the ESO General Director for awarding DDT time to this project, and \rev{to} the ESO staff, in particular Giacomo Beccari and Vincenzo Manieri, for carrying out the observations. We thank Max Pettini, Kevin France, and Simone Scaringi for helpful discussion, and Emma Whelan and Sylvie Cabrit for the UVES observations. We acknowledge with thanks the variable star observations from the AAVSO International Database contributed by observers worldwide and used in this research. During part of the project, SF was funded by an STFC/Isaac Newton Trust studentship in Cambridge. CFM and PCS gratefully acknowledge an ESA Research Fellowship. TJH is funded by an Imperial Junior research fellowship. This work has been supported by the DISCSIM project, grant agreement 341137 funded by the European Research Council under ERC-2013-ADG.
\end{acknowledgements}


\bibliographystyle{aa}
\bibliography{references}

\begin{thebibliography}{64}
\expandafter\ifx\csname natexlab\endcsname\relax\def\natexlab#1{#1}\fi

\bibitem[{{Alcal{\'a}} {et~al.}(2014){Alcal{\'a}}, {Natta}, {Manara}, {Spezzi},
  {Stelzer}, {Frasca}, {Biazzo}, {Covino}, {Randich}, {Rigliaco}, {Testi},
  {Comer{\'o}n}, {Cupani}, \& {D'Elia}}]{2014A&A...561A...2A}
{Alcal{\'a}}, J.~M., {Natta}, A., {Manara}, C.~F., {et~al.} 2014, \aap, 561, A2

\bibitem[{{Alencar} {et~al.}(2005){Alencar}, {Basri}, {Hartmann}, \&
  {Calvet}}]{2005A&A...440..595A}
{Alencar}, S.~H.~P., {Basri}, G., {Hartmann}, L., \& {Calvet}, N. 2005, \aap,
  440, 595

\bibitem[{{Andrews} \& {Williams}(2005)}]{2005ApJ...631.1134A}
{Andrews}, S.~M. \& {Williams}, J.~P. 2005, \apj, 631, 1134

\bibitem[{{Ansdell} {et~al.}(2016){Ansdell}, {Gaidos}, {Williams}, {Kennedy},
  {Wyatt}, {LaCourse}, {Jacobs}, \& {Mann}}]{2016MNRAS.462L.101A}
{Ansdell}, M., {Gaidos}, E., {Williams}, J.~P., {et~al.} 2016, \mnras, 462,
  L101

\bibitem[{{Antipin} {et~al.}(2015){Antipin}, {Belinski}, {Cherepashchuk},
  {Cherjasov}, {Dodin}, {Gorbunov}, {Lamzin}, {Kornilov}, {Kornilov},
  {Potanin}, {Safonov}, {Senik}, {Shatsky}, \&
  {Voziakova}}]{2015IBVS.6126....1A}
{Antipin}, S., {Belinski}, A., {Cherepashchuk}, A., {et~al.} 2015, Information
  Bulletin on Variable Stars, 6126, 1

\bibitem[{{Beristain} {et~al.}(2001){Beristain}, {Edwards}, \&
  {Kwan}}]{2001ApJ...551.1037B}
{Beristain}, G., {Edwards}, S., \& {Kwan}, J. 2001, \apj, 551, 1037

\bibitem[{{Bouvier} {et~al.}(2007){Bouvier}, {Alencar}, {Boutelier},
  {Dougados}, {Balog}, {Grankin}, {Hodgkin}, {Ibrahimov}, {Kun}, {Magakian}, \&
  {Pinte}}]{2007A&A...463.1017B}
{Bouvier}, J., {Alencar}, S.~H.~P., {Boutelier}, T., {et~al.} 2007, \aap, 463,
  1017

\bibitem[{{Bouvier} {et~al.}(2013){Bouvier}, {Grankin}, {Ellerbroek}, {Bouy},
  \& {Barrado}}]{2013A&A...557A..77B}
{Bouvier}, J., {Grankin}, K., {Ellerbroek}, L.~E., {Bouy}, H., \& {Barrado}, D.
  2013, \aap, 557, A77

\bibitem[{{Bozhinova} {et~al.}(2016){Bozhinova}, {Scholz}, {Costigan}, {Lux},
  {Davis}, {Ray}, {Boardman}, {Hay}, {Hewlett}, {Hodos{\'a}n}, \&
  {Morton}}]{2016MNRAS.463.4459B}
{Bozhinova}, I., {Scholz}, A., {Costigan}, G., {et~al.} 2016, \mnras, 463, 4459

\bibitem[{{Cabrit} {et~al.}(2006){Cabrit}, {Pety}, {Pesenti}, \&
  {Dougados}}]{2006A&A...452..897C}
{Cabrit}, S., {Pety}, J., {Pesenti}, N., \& {Dougados}, C. 2006, \aap, 452, 897

\bibitem[{{Casassus} {et~al.}(2015){Casassus}, {Marino}, {P{\'e}rez}, {Roman},
  {Dunhill}, {Armitage}, {Cuadra}, {Wootten}, {van der Plas}, {Cieza}, {Moral},
  {Christiaens}, \& {Montesinos}}]{2015ApJ...811...92C}
{Casassus}, S., {Marino}, S., {P{\'e}rez}, S., {et~al.} 2015, \apj, 811, 92

\bibitem[{{Chou} {et~al.}(2013){Chou}, {Takami}, {Manset}, {Beck}, {Pyo},
  {Chen}, {Panwar}, {Karr}, {Shang}, \& {Liu}}]{2013AJ....145..108C}
{Chou}, M.-Y., {Takami}, M., {Manset}, N., {et~al.} 2013, \aj, 145, 108

\bibitem[{{Clarke} \& {Pringle}(1993)}]{1993MNRAS.261..190C}
{Clarke}, C.~J. \& {Pringle}, J.~E. 1993, \mnras, 261, 190

\bibitem[{{Cox} {et~al.}(2013){Cox}, {Grady}, {Hammel}, {Hornbeck}, {Russell},
  {Sitko}, \& {Woodgate}}]{2013ApJ...762...40C}
{Cox}, A.~W., {Grady}, C.~A., {Hammel}, H.~B., {et~al.} 2013, \apj, 762, 40

\bibitem[{{Crutcher}(1975)}]{1975ApJ...202..634C}
{Crutcher}, R.~M. 1975, \apj, 202, 634

\bibitem[{{Dai} {et~al.}(2015){Dai}, {Facchini}, {Clarke}, \&
  {Haworth}}]{2015MNRAS.449.1996D}
{Dai}, F., {Facchini}, S., {Clarke}, C.~J., \& {Haworth}, T.~J. 2015, \mnras,
  449, 1996

\bibitem[{{Dekker} {et~al.}(2000){Dekker}, {D'Odorico}, {Kaufer}, {Delabre}, \&
  {Kotzlowski}}]{2000SPIE.4008..534D}
{Dekker}, H., {D'Odorico}, S., {Kaufer}, A., {Delabre}, B., \& {Kotzlowski}, H.
  2000, in \procspie, Vol. 4008, Optical and IR Telescope Instrumentation and
  Detectors, ed. M.~{Iye} \& A.~F. {Moorwood}, 534--545

\bibitem[{{Dodin} {et~al.}(2012){Dodin}, {Lamzin}, \&
  {Chuntonov}}]{2012AstL...38..167D}
{Dodin}, A.~V., {Lamzin}, S.~A., \& {Chuntonov}, G.~A. 2012, Astronomy Letters,
  38, 167

\bibitem[{{Dullemond} {et~al.}(2003){Dullemond}, {van den Ancker}, {Acke}, \&
  {van Boekel}}]{2003ApJ...594L..47D}
{Dullemond}, C.~P., {van den Ancker}, M.~E., {Acke}, B., \& {van Boekel}, R.
  2003, \apjl, 594, L47

\bibitem[{{Eisner} {et~al.}(2007){Eisner}, {Hillenbrand}, {White}, {Bloom},
  {Akeson}, \& {Blake}}]{2007ApJ...669.1072E}
{Eisner}, J.~A., {Hillenbrand}, L.~A., {White}, R.~J., {et~al.} 2007, \apj,
  669, 1072

\bibitem[{{Facchini} {et~al.}(2013){Facchini}, {Lodato}, \&
  {Price}}]{2013MNRAS.433.2142F}
{Facchini}, S., {Lodato}, G., \& {Price}, D.~J. 2013, \mnras, 433, 2142

\bibitem[{{Facchini} {et~al.}(2014){Facchini}, {Ricci}, \&
  {Lodato}}]{2014MNRAS.442.3700F}
{Facchini}, S., {Ricci}, L., \& {Lodato}, G. 2014, \mnras, 442, 3700

\bibitem[{{Folha} \& {Emerson}(2001)}]{2001A&A...365...90F}
{Folha}, D.~F.~M. \& {Emerson}, J.~P. 2001, \aap, 365, 90

\bibitem[{{Foucart} \& {Lai}(2011)}]{2011MNRAS.412.2799F}
{Foucart}, F. \& {Lai}, D. 2011, \mnras, 412, 2799

\bibitem[{{France} {et~al.}(2014){France}, {Herczeg}, {McJunkin}, \&
  {Penton}}]{2014ApJ...794..160F}
{France}, K., {Herczeg}, G.~J., {McJunkin}, M., \& {Penton}, S.~V. 2014, \apj,
  794, 160

\bibitem[{{France} {et~al.}(2012){France}, {Schindhelm}, {Herczeg}, {Brown},
  {Abgrall}, {Alexander}, {Bergin}, {Brown}, {Linsky}, {Roueff}, \&
  {Yang}}]{2012ApJ...756..171F}
{France}, K., {Schindhelm}, E., {Herczeg}, G.~J., {et~al.} 2012, \apj, 756, 171

\bibitem[{{Gahm} {et~al.}(1999){Gahm}, {Petrov}, {Duemmler}, {Gameiro}, \&
  {Lago}}]{1999A&A...352L..95G}
{Gahm}, G.~F., {Petrov}, P.~P., {Duemmler}, R., {Gameiro}, J.~F., \& {Lago},
  M.~T.~V.~T. 1999, \aap, 352, L95

\bibitem[{{Ghez} {et~al.}(1997){Ghez}, {White}, \&
  {Simon}}]{1997ApJ...490..353G}
{Ghez}, A.~M., {White}, R.~J., \& {Simon}, M. 1997, \apj, 490, 353

\bibitem[{{Giannini} {et~al.}(2016){Giannini}, {Lorenzetti}, {Harutyunyan}, {Li
  Causi}, {Antoniucci}, {Arkharov}, {Larionov}, {Strafella}, {Carini}, {Di
  Paola}, \& {Speziali}}]{2016A&A...588A..20G}
{Giannini}, T., {Lorenzetti}, D., {Harutyunyan}, A., {et~al.} 2016, \aap, 588,
  A20

\bibitem[{{Grankin} {et~al.}(2007){Grankin}, {Melnikov}, {Bouvier}, {Herbst},
  \& {Shevchenko}}]{2007A&A...461..183G}
{Grankin}, K.~N., {Melnikov}, S.~Y., {Bouvier}, J., {Herbst}, W., \&
  {Shevchenko}, V.~S. 2007, \aap, 461, 183

\bibitem[{{Gregory} {et~al.}(2012){Gregory}, {Donati}, {Morin}, {Hussain},
  {Mayne}, {Hillenbrand}, \& {Jardine}}]{2012ApJ...755...97G}
{Gregory}, S.~G., {Donati}, J.-F., {Morin}, J., {et~al.} 2012, \apj, 755, 97

\bibitem[{{Grinin} {et~al.}(1998){Grinin}, {Rostopchina}, \&
  {Shakhovskoi}}]{1998AstL...24..802G}
{Grinin}, V.~P., {Rostopchina}, A.~N., \& {Shakhovskoi}, D.~N. 1998, Astronomy
  Letters, 24, 802

\bibitem[{{Hartigan} {et~al.}(1995){Hartigan}, {Edwards}, \&
  {Ghandour}}]{1995ApJ...452..736H}
{Hartigan}, P., {Edwards}, S., \& {Ghandour}, L. 1995, \apj, 452, 736

\bibitem[{{Ingleby} {et~al.}(2013){Ingleby}, {Calvet}, {Herczeg}, {Blaty},
  {Walter}, {Ardila}, {Alexander}, {Edwards}, {Espaillat}, {Gregory},
  {Hillenbrand}, \& {Brown}}]{2013ApJ...767..112I}
{Ingleby}, L., {Calvet}, N., {Herczeg}, G., {et~al.} 2013, \apj, 767, 112

\bibitem[{{Johnson} \& {Morgan}(1953)}]{1953ApJ...117..313J}
{Johnson}, H.~L. \& {Morgan}, W.~W. 1953, \apj, 117, 313

\bibitem[{Kafka(2016)}]{aavso}
Kafka, S. 2016, Observations from the AAVSO International Database,
  http://www.aavso.org

\bibitem[{{Kwan} \& {Fischer}(2011)}]{2011MNRAS.411.2383K}
{Kwan}, J. \& {Fischer}, W. 2011, \mnras, 411, 2383

\bibitem[{{L{\'o}pez-Mart{\'{\i}}n} {et~al.}(2003){L{\'o}pez-Mart{\'{\i}}n},
  {Cabrit}, \& {Dougados}}]{2003A&A...405L...1L}
{L{\'o}pez-Mart{\'{\i}}n}, L., {Cabrit}, S., \& {Dougados}, C. 2003, \aap, 405,
  L1

\bibitem[{{Manara} {et~al.}(2013){Manara}, {Beccari}, {Da Rio}, {De Marchi},
  {Natta}, {Ricci}, {Robberto}, \& {Testi}}]{2013A&A...558A.114M}
{Manara}, C.~F., {Beccari}, G., {Da Rio}, N., {et~al.} 2013, \aap, 558, A114

\bibitem[{{Marino} {et~al.}(2015){Marino}, {Perez}, \&
  {Casassus}}]{2015ApJ...798L..44M}
{Marino}, S., {Perez}, S., \& {Casassus}, S. 2015, \apjl, 798, L44

\bibitem[{{Mathis} {et~al.}(1977){Mathis}, {Rumpl}, \&
  {Nordsieck}}]{1977ApJ...217..425M}
{Mathis}, J.~S., {Rumpl}, W., \& {Nordsieck}, K.~H. 1977, \apj, 217, 425

\bibitem[{{McJunkin} {et~al.}(2013){McJunkin}, {France}, {Burgh}, {Herczeg},
  {Schindhelm}, {Brown}, \& {Brown}}]{2013ApJ...766...12M}
{McJunkin}, M., {France}, K., {Burgh}, E.~B., {et~al.} 2013, \apj, 766, 12

\bibitem[{{Modigliani} {et~al.}(2010){Modigliani}, {Goldoni}, {Royer},
  {Haigron}, {Guglielmi}, {Fran{\c c}ois}, {Horrobin}, {Bristow}, {Vernet},
  {Moehler}, {Kerber}, {Ballester}, {Mason}, \&
  {Christensen}}]{2010SPIE.7737E..28M}
{Modigliani}, A., {Goldoni}, P., {Royer}, F., {et~al.} 2010, in SPIE Conference
  Series, Vol. 7737, SPIE Conference Series, 28

\bibitem[{{Morton} \& {Smith}(1973)}]{1973ApJS...26..333M}
{Morton}, D.~C. \& {Smith}, W.~H. 1973, \apjs, 26, 333

\bibitem[{{Ostriker}(1994)}]{1994ApJ...424..292O}
{Ostriker}, E.~C. 1994, \apj, 424, 292

\bibitem[{{Pascucci} {et~al.}(2015){Pascucci}, {Edwards}, {Heyer}, {Rigliaco},
  {Hillenbrand}, {Gorti}, {Hollenbach}, \& {Simon}}]{2015ApJ...814...14P}
{Pascucci}, I., {Edwards}, S., {Heyer}, M., {et~al.} 2015, \apj, 814, 14

\bibitem[{{Petrov} {et~al.}(2015){Petrov}, {Gahm}, {Djupvik}, {Babina},
  {Artemenko}, \& {Grankin}}]{2015A&A...577A..73P}
{Petrov}, P.~P., {Gahm}, G.~F., {Djupvik}, A.~A., {et~al.} 2015, \aap, 577, A73

\bibitem[{{Petrov} {et~al.}(2001{\natexlab{a}}){Petrov}, {Gahm}, {Gameiro},
  {Duemmler}, {Ilyin}, {Laakkonen}, {Lago}, \&
  {Tuominen}}]{2001A&A...369..993P}
{Petrov}, P.~P., {Gahm}, G.~F., {Gameiro}, J.~F., {et~al.} 2001{\natexlab{a}},
  \aap, 369, 993

\bibitem[{{Petrov} {et~al.}(2001{\natexlab{b}}){Petrov}, {Pelt}, \&
  {Tuominen}}]{2001A&A...375..977P}
{Petrov}, P.~P., {Pelt}, J., \& {Tuominen}, I. 2001{\natexlab{b}}, \aap, 375,
  977

\bibitem[{{Rodgers} {et~al.}(2002){Rodgers}, {Wooden}, {Grinin}, {Shakhovsky},
  \& {Natta}}]{2002ApJ...564..405R}
{Rodgers}, B., {Wooden}, D.~H., {Grinin}, V., {Shakhovsky}, D., \& {Natta}, A.
  2002, \apj, 564, 405

\bibitem[{{Rodriguez} {et~al.}(2013){Rodriguez}, {Pepper}, {Stassun}, {Siverd},
  {Cargile}, {Beatty}, \& {Gaudi}}]{2013AJ....146..112R}
{Rodriguez}, J.~E., {Pepper}, J., {Stassun}, K.~G., {et~al.} 2013, \aj, 146,
  112

\bibitem[{{Rodriguez} {et~al.}(2015){Rodriguez}, {Pepper}, {Stassun}, {Siverd},
  {Cargile}, {Weintraub}, {Beatty}, {Gaudi}, {Mamajek}, \&
  {Sanchez}}]{2015AJ....150...32R}
{Rodriguez}, J.~E., {Pepper}, J., {Stassun}, K.~G., {et~al.} 2015, \aj, 150, 32

\bibitem[{{Rodriguez} {et~al.}(2016){Rodriguez}, {Reed}, {Siverd}, {Pepper},
  {Stassun}, {Gaudi}, {Weintraub}, {Beatty}, {Lund}, \&
  {Stevens}}]{2016AJ....151...29R}
{Rodriguez}, J.~E., {Reed}, P.~A., {Siverd}, R.~J., {et~al.} 2016, \aj, 151, 29

\bibitem[{{Scaringi} {et~al.}(2016){Scaringi}, {Manara}, {Barenfeld}, {Groot},
  {Isella}, {Kenworthy}, {Knigge}, {Maccarone}, {Ricci}, \&
  {Ansdell}}]{2016MNRAS.463.2265S}
{Scaringi}, S., {Manara}, C.~F., {Barenfeld}, S.~A., {et~al.} 2016, \mnras,
  463, 2265

\bibitem[{{Schneider} {et~al.}(2015{\natexlab{a}}){Schneider}, {France},
  {G{\"u}nther}, {Herczeg}, {Robrade}, {Bouvier}, {McJunkin}, \&
  {Schmitt}}]{2015A&A...584A..51S}
{Schneider}, P.~C., {France}, K., {G{\"u}nther}, H.~M., {et~al.}
  2015{\natexlab{a}}, \aap, 584, A51

\bibitem[{{Schneider} {et~al.}(2015{\natexlab{b}}){Schneider}, {G{\"u}nther},
  {Robrade}, {Facchini}, {Hodapp}, {Manara}, {Perdelwitz}, {Schmitt},
  {Skinner}, \& {Wolk}}]{2015A&A...584L...9S}
{Schneider}, P.~C., {G{\"u}nther}, H.~M., {Robrade}, J., {et~al.}
  2015{\natexlab{b}}, \aap, 584, L9

\bibitem[{{Shenavrin} {et~al.}(2015){Shenavrin}, {Petrov}, \&
  {Grankin}}]{2015IBVS.6143....1S}
{Shenavrin}, V.~I., {Petrov}, P.~P., \& {Grankin}, K.~N. 2015, Information
  Bulletin on Variable Stars, 6143, 1

\bibitem[{{Stepanovs} {et~al.}(2014){Stepanovs}, {Fendt}, \&
  {Sheikhnezami}}]{2014ApJ...796...29S}
{Stepanovs}, D., {Fendt}, C., \& {Sheikhnezami}, S. 2014, \apj, 796, 29

\bibitem[{{Takami} {et~al.}(2016){Takami}, {Wei}, {Chou}, {Karr}, {Beck},
  {Manset}, {Chen}, {Kurosawa}, {Fukagawa}, {White}, {Galv{\'a}n-Madrid},
  {Liu}, {Pyo}, \& {Donati}}]{2016ApJ...820..139T}
{Takami}, M., {Wei}, Y.-J., {Chou}, M.-Y., {et~al.} 2016, \apj, 820, 139

\bibitem[{{van Leeuwen}(2007)}]{2007A&A...474..653V}
{van Leeuwen}, F. 2007, \aap, 474, 653

\bibitem[{{Vernet} {et~al.}(2011){Vernet}, {Dekker}, {D'Odorico}, {Kaper},
  {Kjaergaard}, {Hammer}, {Randich}, {Zerbi}, {Groot}, {Hjorth}, {Guinouard},
  {Navarro}, {Adolfse}, {Albers}, {Amans}, {Andersen}, {Andersen}, {Binetruy},
  {Bristow}, {Castillo}, {Chemla}, {Christensen}, {Conconi}, {Conzelmann},
  {Dam}, {de Caprio}, {de Ugarte Postigo}, {Delabre}, {di Marcantonio},
  {Downing}, {Elswijk}, {Finger}, {Fischer}, {Flores}, {Fran{\c c}ois},
  {Goldoni}, {Guglielmi}, {Haigron}, {Hanenburg}, {Hendriks}, {Horrobin},
  {Horville}, {Jessen}, {Kerber}, {Kern}, {Kiekebusch}, {Kleszcz}, {Klougart},
  {Kragt}, {Larsen}, {Lizon}, {Lucuix}, {Mainieri}, {Manuputy}, {Martayan},
  {Mason}, {Mazzoleni}, {Michaelsen}, {Modigliani}, {Moehler}, {M{\o}ller},
  {Norup S{\o}rensen}, {N{\o}rregaard}, {P{\'e}roux}, {Patat}, {Pena}, {Pragt},
  {Reinero}, {Rigal}, {Riva}, {Roelfsema}, {Royer}, {Sacco}, {Santin},
  {Schoenmaker}, {Spano}, {Sweers}, {Ter Horst}, {Tintori}, {Tromp}, {van
  Dael}, {van der Vliet}, {Venema}, {Vidali}, {Vinther}, {Vola}, {Winters},
  {Wistisen}, {Wulterkens}, \& {Zacchei}}]{2011A&A...536A.105V}
{Vernet}, J., {Dekker}, H., {D'Odorico}, S., {et~al.} 2011, \aap, 536, A105

\bibitem[{{Vink} {et~al.}(2005){Vink}, {Drew}, {Harries}, {Oudmaijer}, \&
  {Unruh}}]{2005MNRAS.359.1049V}
{Vink}, J.~S., {Drew}, J.~E., {Harries}, T.~J., {Oudmaijer}, R.~D., \& {Unruh},
  Y. 2005, \mnras, 359, 1049

\bibitem[{{White} \& {Ghez}(2001)}]{2001ApJ...556..265W}
{White}, R.~J. \& {Ghez}, A.~M. 2001, \apj, 556, 265

\bibitem[{{Woitas} {et~al.}(2001){Woitas}, {Leinert}, \&
  {K{\"o}hler}}]{2001A&A...376..982W}
{Woitas}, J., {Leinert}, C., \& {K{\"o}hler}, R. 2001, \aap, 376, 982

\end{thebibliography}

\end{document}